\documentclass[aps,pre,groupedaddress]{revtex4-1}

\usepackage{graphicx}
\usepackage{bm}

\def\varstackrel#1#2{\mathrel{\mathop{#1}\limits_{#2}}}
\newcommand{\Tr}{\mbox{Tr}~}
\newcommand{\Trs}[1]{\varstackrel{\Tr}{\small #1}}

\usepackage{color}	
\usepackage{ulem}

\graphicspath{{./fig/}}

\begin{document}

\title{Statistical mechanical evaluation of spread spectrum watermarking
model with image restoration}

\author{Masaki Kawamura} \email[]{kawamura@sci.yamaguchi-u.ac.jp}
\affiliation{Graduate School of Sciences and Technology for Innovation, Yamaguchi University \\
Yoshida 1677-1, Yamaguchi, 753-8512 Japan}

\author{Kao Hayashi}
\author{Tatsuya Uezu}
\affiliation{Graduate School of Humanities and Sciences, Nara Women's University
\\ Kitauoyanishi-machi, Nara 630-8506 Japan}

\author{Masato Okada}
\affiliation{Graduate School of Frontier Sciences, The University of Tokyo\\
 Kashiwanoha 5-1-5, Kashiwa, 277-8561 Japan}

\date{\today}
\begin{abstract}
In cases in which an original image is blind, a decoding method where
both the image and the messages can be estimated simultaneously is
desirable. We propose a spread spectrum watermarking model with image
restoration based on Bayes estimation. We therefore need to assume some
prior probabilities. The probability for estimating the messages is
given by the uniform distribution, and the ones for the image are given
by the infinite range model and 2D Ising model. Any attacks from
unauthorized users can be represented by channel models. We can obtain
the estimated messages and image by maximizing the posterior
probability.

We analyzed the performance of the proposed method by the replica method
in the case of the infinite range model. We first calculated the
theoretical values of the bit error rate from obtained saddle point
equations and then verified them by computer simulations. For this
purpose, we assumed that the image is binary and is generated from a
given prior probability. We also assume that attacks can be represented
by the Gaussian channel. The computer simulation retults agreed with the
theoretical values.

In the case of prior probability given by the 2D Ising model,
in which each pixel is statically connected with four-neighbors,
we evaluated the decoding performance by computer simulations,
since the replica theory could not be applied. 
Results using the 2D Ising model
showed that the proposed method with image restoration is as effective
as the infinite range model for decoding messages.

We compared the performances in a case in which the image was blind and
one in which it was informed. The difference between these cases was
small as long as the embedding and attack rates were small.  This
demonstrates that the proposed method with simultaneous estimation is
effective as a watermarking decoder.
\end{abstract}



\published{June 26, 2019}

\maketitle

\section{Introduction}

Digital watermarking is attracting attention for its potential
application against the misuse of digital content. The basic idea of
digital watermarking is that some hidden messages or watermarks such as
a copyright or user ID are invisibly embedded in digital cover
content. For image watermarking, we need to pay attention to both the
hidden messages and the images themselves. Either watermarks are simply
embedded by adding them to the cover content \cite{Cox1996,CoxBook}, or
the cover content is transformed by discrete cosine transform (DCT)
\cite{Cox1997} or wavelet transform \cite{OhnishiMatsui1996} and the
watermarks are embedded in the transform domain.
For the watermarks themselves, random binary bit or Gaussian sequences
are usually used for the embedding \cite{Cox1996,CoxBook,Cox1997}.  The
messages may be encoded \cite{Cox1999}. The spectrum spreading method is
an efficient, robust method. In this paper, we consider a decoding
algorithm for the spectrum spreading method.

The basic spectrum spreading technique is also used in code division
nultiple access (CDMA) \cite{Verdu1989}, where multiple users can
transmit their information at the same time and within the same
cell. Multiuser interference needs to be considered for the
CDMA multiuser demodulator problem. Recently Bayes optimum solutions
have been proposed on statistical mechanics
\cite{TTanaka2001,TTanaka2002,Nishimori2001,YoshidaUezu2007}.
In spread spectrum digital watermarking
\cite{Cox1996,Cox1997,CoxBook}, watermarks are generated by
spreading the messages. Stego images, which are marked images, are
generated by embedding these watermarks in the original images. 
Attacks to or misuses of the stego images can be represented by channel
models. We must estimate the hidden messages from tampered images while
reducing multi-watermarks interference. 

In an informed case -- that is, a case in which the original image is
known to the decoder -- we can determine the difference between the
original and the tampered images.
Using a framework of the Bayes estimation
\cite{TTanaka2001,TTanaka2002,Nishimori2001}, we can estimate these
messages from the received messages by maximizing the posterior
probability \cite{SendaKawamura2009}.
In contrast, in the blind case -- that is, a case in which the original
image is unknown -- we need to estimate the original images from the
tampered images.  Watermarks are treated as noises against the image,
and therefore, image estimation need to be applied to such a
case. Assuming the prior probability of images, we introduce Bayes image
estimation
\cite{GemanGeman1984,NishimoriWong1999,Inoue2001,KTanaka2002JPA,Nishimori2001}
to the blind watermarking model.
In order to estimate original images, we must assume the model used to
generate the images.
Natural images are usually represented as 8 bits per
pixel. Using the least significant bit (LSB) or parity of the natural
images, binary images can easily be generated. Embedding the watermarks
into the binary images is now common \cite{Fridrich2005}. In this
paper, we use binary images.

Performance of the blind digital watermarking model has not yet been
sufficiently evaluated. We therefore evaluate the average performance of
this model. In particular, in the blind case, we propose a method in
which both messages and the original image can be estimated at the same
time.  In order to evaluate the proposed method, we derive saddle point
equations by the replica method and then calculate the theoretical bit
error rate.
For the theoretical evaluation using the replica method, we assume the
infinite range model as prior probability of images. Moreover, we
evaluate the case of the 2D Ising model as a prior probability by
computer simulations.

\begin{figure*}[tb]
 \hfill\includegraphics[width=50mm]{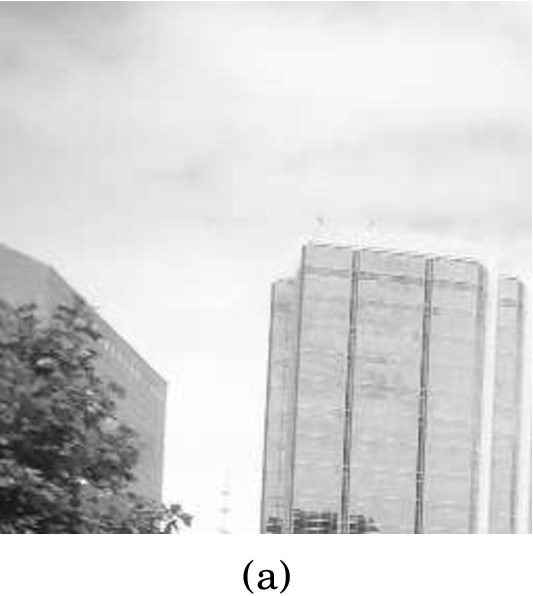}
 \hfill\includegraphics[width=50mm]{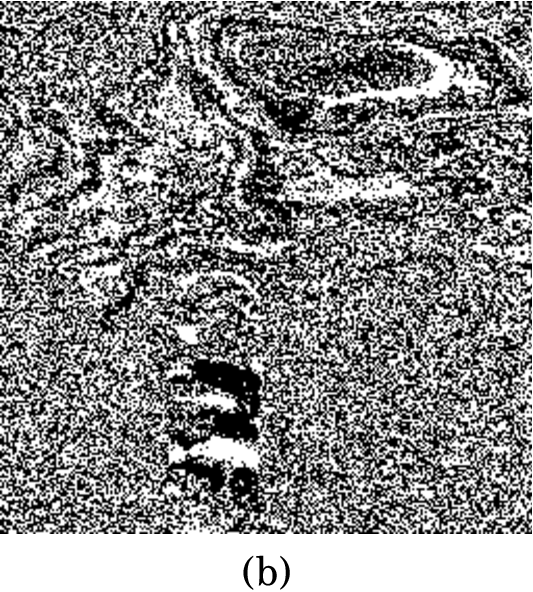}
 \hfill\includegraphics[width=50mm]{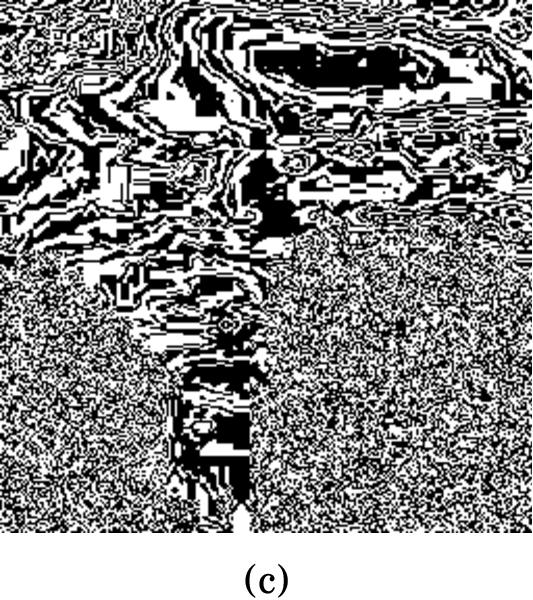}
 \hfill\mbox{}

 \caption{Sample of natural and parity images, (a) original image,
 (b) parity of uncompressed image, and (c) parity of JPEG image. The
 parity images are generated from parity bits. }
 \label{fig:IHeval03}
\end{figure*}

Now, we discuss the feasibility of representing original images by the
infinite range model and 2D Ising model. Watermarking methods such as
the wet paper code \cite{Fridrich2005} and matrix embedding
\cite{Fridrich2006} methods assume that content consists of binary
data. Specifically, the original images to be embedded are generated by
calculating LSB or parity bits.
We refer to a binary image consisting of parity bits as a parity image.
Figure~\ref{fig:IHeval03} shows the parity images generated from a
natural image, where (a) is the original natural image and (b) shows the
parity image from the uncompressed natural image of (a). The parity
image in (c) is generated after JPEG compression of (a). 
The black and white pixels represent the parity bits $0$ and $1$,
respectively. 
Figuratively speaking, from these images, we can find that part of the
parity images (b) and (c) can be seen as an image generated from the
infinite range model and other part with some clusters can be
seen as one from the 2D Ising model.
Since we can evaluate our method in
theory, it is reasonable to introduce some image generation models.

The rest of this paper is organized as follows. 
Section \ref{sec:intro} gives an overview of our watermarking model. We explain that
both messages and images can be estimated by maximizing the posterior
probability. Section \ref{sec:BER} describes the saddle point equations
derived by the replica method in order to evaluate our method.  Section
\ref{sec:sim} shows the results obtained by theory and computer simulations.
We conclude the paper in Section \ref{sec:conc}.

\section{digital watermarking model} \label{sec:intro}

We describe a basic watermarking model in an informed case and an
image restoration model before proposing our blind watermarking model.

\subsection{Informed case}

When a decoder has been informed of an original image, the informed
spread spectrum watermarking model can correspond to the CDMA model.
$K$-bit messages $\boldsymbol{s}=(s_1,s_2,\cdots,s_K)^{\top}$ are embedded in
an original image in layers, where $s_i=\pm1$. We assume the prior
probability of messages is a uniform distribution given by
\begin{equation}
 P\left(\boldsymbol{s}\right) = \frac{1}{2^K} .
  \label{eqn:apriori_s}
\end{equation}
Each message $s_i$ is spread by a specific spreading code
$\boldsymbol{\xi}_i=(\xi^{1}_i,\xi^{2}_i,\cdots,\xi^{N}_i)^{\top}$, and
watermarks are obtained by summing the $K$ spread messages. The length
of the spread codes -- that is, the chip rate -- is equal to the size of
the image, $N$. Each element of spreading codes $\xi^{\mu}_i$
takes $\pm1$ with probability
\begin{equation}
 P\left(\xi^{\mu}_i=\pm1\right)=\frac{1}{2} .
\end{equation}
Here, $(\xi^{\mu}_i)^2=1$. 
$\mu$-th watermark $w_{\mu}$ is represented by 
\begin{eqnarray}
 w_{\mu} &=& \frac{1}{\sqrt{K}}\sum_{i=1}^{K} \xi^{\mu}_i s_i 
  \;,\;\;\mu=1,2,\cdots,N .
\end{eqnarray}
The stego image or marked image $\boldsymbol{X}$ is created by
adding the watermark $\boldsymbol{w}$ to the original image
$\boldsymbol{f}$; that is, $X_{\mu} = f_{\mu} + w_{\mu}$.
We ignore any embedding errors, because they are almost always small
enough to be negligent.

Here, assume we have received a tampered stego image that is attacked by
an illegal user. We can consider this attack the deterioration process
of an image.  Attacks can be represented as noise in the communication
channel \cite{Cox1999,HernandezPerezGonzalez1999,Su1999}.  We assume the
channel is represented by the additive white Gaussian noise (AWGN)
channel. Therefore, the conditional probability of the tampered image
$\boldsymbol{r}$ given messages $\boldsymbol{s}$ is given by
\begin{eqnarray}
 P\left(\boldsymbol{r}\vert \boldsymbol{s}\right) 
  =\prod_{\mu=1}^{N} P\left(r_{\mu}\vert \boldsymbol{s}\right)  
  \propto
  \exp\left[-\frac{1}{2\sigma_0^2}
       \sum_{\mu=1}^N \left(r_{\mu}-w_{\mu} \right)^2 \right] ,
\end{eqnarray}
where noise obeys the Gaussian distribution ${\cal N}(0,\sigma^2_0)$.

What we want to know is how many messages the decoder can retrieve from
the tampered image. We therefore need to estimate messages
$\boldsymbol{s}$ and then calculate the bit error rate.
In order to estimate the messages, the posterior probability of messages
$\boldsymbol{s}$ given the tampered image $\boldsymbol{r}$ should be
computed. Since the true parameter $\sigma_0^2$ is unknown, we set a
parameter as $\sigma^2$. From (\ref{eqn:apriori_s}) and Bayes theorem, the 
posterior probability is given by
\begin{eqnarray}
 P\left(\boldsymbol{s}\vert \boldsymbol{r}\right) 
  &=& \frac{P\left(\boldsymbol{r}\vert \boldsymbol{s}\right) 
  P\left(\boldsymbol{s}\right)}
  {\sum_{\boldsymbol{s}}
  P\left(\boldsymbol{r}\vert \boldsymbol{s}\right) 
  P\left(\boldsymbol{s}\right) }  \\
 &=& \frac{1}{Z} 
  \exp\left[-\frac{1}{2\sigma^2}
   \sum_{\mu=1}^N \left(r_{\mu}-w_{\mu}\right)^2 \right] ,
  \label{eqn:post_s} \\
 Z &=& \Trs{\boldsymbol{s}}
 \exp\left[-\frac{1}{2\sigma^2}
      \sum_{\mu=1}^N \left(r_{\mu}-w_{\mu}\right)^2 \right] ,
\end{eqnarray}
where $Z$ is a normalization factor called a partition function.
The watermark $w_{\mu}$ is a function of the messages $\boldsymbol{s}$.
$\Trs{s}$ stands for the summation over $s$.

For a maximum a posteriori (MAP) estimation, 
the estimated messages $\boldsymbol{\widehat{s}}$ are given by
\begin{eqnarray}
 \boldsymbol{\widehat{s}} &=&
  \mathrm{arg}\mathrel{\mathop{\max}\limits_{\small \boldsymbol{x}}}
  P\left(\boldsymbol{x}\vert \boldsymbol{r}\right) ,
\end{eqnarray}
where $\boldsymbol{x}=(x_1,x_2,\dots,x_K)^{\top}$ are variables that
represent messages. 
For a maximum posterior marginal (MPM) estimation, 
the estimated messages $\boldsymbol{\widehat{s}}$ are given by
\begin{eqnarray}
 \widehat{s}_i &=&
  \mathrm{arg}\mathrel{\mathop{\max}\limits_{\small x_i}}
  \sum_{\boldsymbol{x}\backslash x_i}
  P\left(\boldsymbol{x}\vert \boldsymbol{r}\right) ,
\end{eqnarray}
where summation $\sum_{\boldsymbol{x}\setminus x_i}$ is a summation over
${\boldsymbol x}$ excepting $x_i$.
With that, we can obtain a Bayes optimum estimation.

\subsection{Image restoration model}

\begin{figure}[tb]
 \hfill\includegraphics[width=50mm]{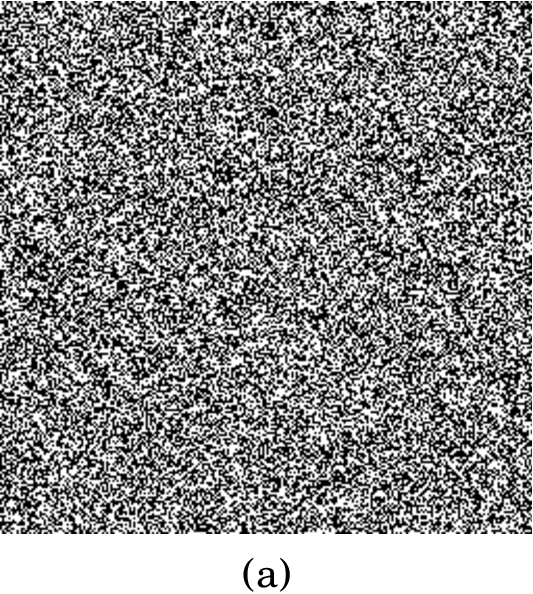}
 \hfill\includegraphics[width=50mm]{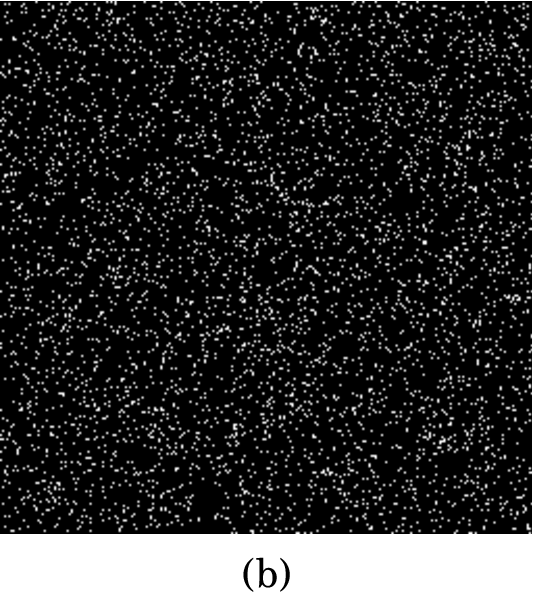}
 \hfill\includegraphics[width=50mm]{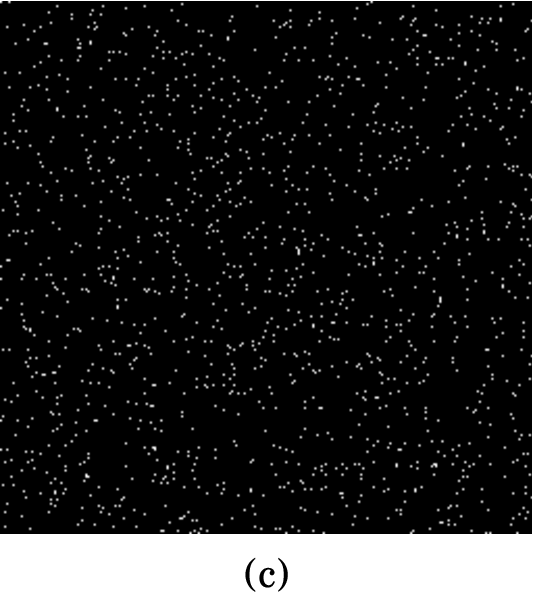}
 \hfill\mbox{}

 \caption{Images generated by infinite range model ($256\times256$
 pixels) with smooth parameters (a)~$\alpha_0=1.0$, (b)~$\alpha_0=1.5$, and
 (c)~$\alpha_0=2.0$.}
 \label{fig:IRMSample}
\end{figure}

It is difficult to formulate natural images. In the image restoration
method based on Bayes estimation, the original images are assumed to be
generated from some probability distribution
\cite{GemanGeman1984,NishimoriWong1999,KTanaka2002JPA}.
In this paper, we assume that the original images consist of $N$ pixels and 
that the pixels are binary
\cite{GemanGeman1984,NishimoriWong1999,KTanaka2002JPA}.
Moreover, we consider the infinite range model \cite{Nishimori2001}
and the 2D Ising model as image generating models.
The prior probability of the infinite range model is given by 
\begin{eqnarray}
 P\left(\boldsymbol{f}\right) 
  &\propto& \exp\left[\frac{\alpha_0}{N}
                 \sum_{\mu<\nu}f_{\mu}f_{\nu}\right], 
  \label{eqn:apriori_f}
\end{eqnarray}
where parameter $\alpha_0$ represents the smoothness of an image
and the summation $\sum_{\mu<\nu}$ runs over all pairs of different
indexes $\mu,\nu$. 
Figure~\ref{fig:IRMSample} shows some images generated by the infinite
range model with $\alpha_0=1.0, 1.5$, and $2.0$.  
Although sites in the infinite range model 
are not intrinsically lined up, we arrange these sites on the
two-dimensional lattice like $256\times256$ pixel images. 
In the case of (a),
the image looks like high-frequency snow noise, while with the larger
$\alpha_0$ in (c), smooth images appear.

For a while, leaving the watermarking scheme aside, we concentrate
exclusively on image restoration from a tampered image. In fact,
the embedding process of the watermarks can be considered a Gaussian
channel.  Therefore, we assume the deterioration process from the
original image to the tampered image is a Gaussian channel.  In this case,
the probability of the tampered image $\boldsymbol{r}$ given the
original image $\boldsymbol{f}$ is given by
\begin{eqnarray}
 P\left(\boldsymbol{r}\vert \boldsymbol{f}\right) 
  = \! \prod_{\mu=1}^{N} P\left(r_{\mu}\vert \boldsymbol{f}\right) 
  \propto
  \exp\left[-\frac{1}{2\sigma_0^2}
       \sum_{\mu=1}^N \left(r_{\mu}-f_{\mu} \right)^2 \right] .
\end{eqnarray}
For the infinite range model, from Bayes theorem, the image maximizing
the posterior probability,
\begin{eqnarray}
 P\left(\boldsymbol{f}\vert \boldsymbol{r}\right) 
 &=& \frac{1}{Z} 
  \exp\left[-\frac{1}{2\sigma^2}
   \sum_{\mu=1}^N \left(r_{\mu}-f_{\mu}\right)^2  
   +\frac{\alpha}{N}\sum_{\mu<\nu}f_{\mu}f_{\nu} \right] ,
   \label{eqn:post_f} \\
 Z &=& 
 \Trs{\boldsymbol{f}}
 P\left(\boldsymbol{r}\vert \boldsymbol{f}\right) 
 P\left(\boldsymbol{f}\right), 
\end{eqnarray}
can be chosen as the estimation image. Since the true parameters
$\sigma_0^2, \alpha_0$ are unknown, parameters $\sigma^2$ and $\alpha$
are used.

\subsection{Blind case}

When the original image is unknown or blind at the decoder, 
both the messages and the image should be estimated at the same time.  This
method requires the posterior probability of messages $\boldsymbol{s}$
and image $\boldsymbol{f}$ given the tampered image $\boldsymbol{r}$.
Since the probability of the tampered image $\boldsymbol{r}$ is given by
\begin{eqnarray}
 P\left(\boldsymbol{r}\vert \boldsymbol{s},\boldsymbol{f}\right) 
  &\propto&
  \exp\left[-\frac{1}{2\sigma_0^2}
       \sum_{\mu=1}^N \left(r_{\mu}-w_{\mu}-f_{\mu} \right)^2 \right] ,
\end{eqnarray}
and the prior probabilities are given by (\ref{eqn:apriori_s}) and
(\ref{eqn:apriori_f}), the posterior probability can be given by
\begin{eqnarray}
 P\left(\boldsymbol{s},\boldsymbol{f}\vert \boldsymbol{r}\right) 
  &=& \frac{P\left(\boldsymbol{r}\vert \boldsymbol{s},\boldsymbol{f}\right) 
  P\left(\boldsymbol{s}\right) P\left(\boldsymbol{f}\right)}
  {\sum_{\boldsymbol{s},\boldsymbol{f}}
  P\left(\boldsymbol{r}\vert \boldsymbol{s},\boldsymbol{f}\right) 
  P\left(\boldsymbol{s}\right) P\left(\boldsymbol{f}\right)}  \\
 &=& \frac{1}{Z} 
  \exp\left[-\frac{1}{2\sigma^2}
   \sum_{\mu=1}^N \left(r_{\mu}-w_{\mu}-f_{\mu}\right)^2  
   +\frac{\alpha}{N}\sum_{\mu<\nu}f_{\mu}f_{\nu} \right] \!,
   \label{eqn:post_sf} 
\end{eqnarray}
where 
\begin{eqnarray}
 Z &=& 
 \Trs{\boldsymbol{s}, \boldsymbol{f}}
 P\left(\boldsymbol{r}\vert \boldsymbol{s},\boldsymbol{f}\right) 
 P\left(\boldsymbol{f}\right) .
\end{eqnarray}
Constant $P\left(\boldsymbol{s}\right)$ is reducible.
Since the true parameters
$\sigma_0^2$ and $\alpha_0$ are unknown, parameters $\sigma^2$ and $\alpha$
are used.
Now, we rewrite the posterior probability in a different form using the
Hamiltonian $H(\boldsymbol{s},\boldsymbol{f})$ as
$P(\boldsymbol{s},\boldsymbol{f}\vert
\boldsymbol{r})=\exp(-H(\boldsymbol{s},\boldsymbol{f})/\sigma^2)/Z$. 
We can then obtain the Hamiltonian, 
\begin{eqnarray}
 H\left(\boldsymbol{s},\boldsymbol{f} \right)
  &=& \frac{1}{2\beta}\sum_{i=1}^{K}\sum_{j=1}^{K}J_{ij}s_is_j
  -\frac{1}{\sqrt{\beta}}\sum_{i=1}^{K}h_is_i
  -\sum_{\mu=1}^{N}r_{\mu}f_{\mu} 
   \nonumber \\ 
  &+& \frac{1}{\sqrt{K}}\sum_{\mu=1}^{N}\sum_{i=1}^{K}f_{\mu}\xi_i^{\mu}s_i 
  -\frac{\alpha\sigma^2}{N} \sum_{\mu<\nu} 
  f_{\mu}f_{\nu} ,
  \label{eqn:Hamiltonian}
\end{eqnarray}
where $\beta$ stands for embedding rate $\beta=K/N$, and 
\begin{eqnarray}
 J_{ij} &=& \frac{1}{N}\sum_{\mu=1}^{N} \xi_{i}^{\mu} \xi_{j}^{\mu} \;,\;\;
 h_i = \frac{1}{\sqrt{N}}\sum_{\mu=1}^{N} \xi_{i}^{\mu}r_{\mu} .
  \label{def_Jij_hi} 
\end{eqnarray}
From MAP and MPM estimations, the estimated messages
$\boldsymbol{\widehat{s}}$ and estimated image
$\boldsymbol{\widehat{f}}$ are given by
\begin{eqnarray}
 \mathrm{MAP} &:& \left(\boldsymbol{\widehat{s},\widehat{f}}\right) =
  \mathrm{arg}\mathrel{\mathop{\max}\limits_{\small (\boldsymbol{x,g})}}
  P\left(\boldsymbol{x},\boldsymbol{g}\vert \boldsymbol{r}\right) , \\
 \mathrm{MPM} &:&
 \widehat{S}_i =
  \mathrm{arg}\mathrel{\mathop{\max}\limits_{\small X_i}}
  \sum_{\boldsymbol{X}\backslash X_i}
  P\left(\boldsymbol{X}\vert \boldsymbol{r}\right) ,
\end{eqnarray}
where $\boldsymbol{x}=(x_1,x_2,\dots,x_K)^{\top}$ and
$\boldsymbol{g}=(g_1,g_2,\cdots,g_N)^{\top}$ stand for variables of
messages and image, respectively.  For MPM estimation,
$\boldsymbol{S}=(s_1,\cdots,s_K, f_1,\cdots,f_N)^{\top}$ stand for the
true values of original messages and image.
$\widehat{S}_i$ represents each element of the estimated  messages
$\boldsymbol{\widehat{s}}$ and image $\boldsymbol{\widehat{f}}$,
that is, $\widehat{S}_i\in\left\{\widehat{s}_1,\cdots,\widehat{s}_K,
\widehat{f}_1,\cdots,\widehat{f}_N\right\}$. 
$\boldsymbol{X}=(x_1,x_2,\dots,x_K, g_1,g_2,\cdots,g_N)^{\top}$
represents the corresponding variables of messages and image,
and $X_i$ is $i$-th element in $\boldsymbol{X}$ corresponding to
$\widehat{S}_i$.
The MPM estimation for the CDMA model can be seen in
\cite{Nishimori2001}.

\section{theoretical evaluation} \label{sec:BER}

\subsection{Bit error rate}

The accuracy for estimated messages can be measured by bit error rate
($\mathrm{BER}$), as
\begin{eqnarray}
 \mathrm{BER}_m &=& \frac{1-d_m}{2}, 
\end{eqnarray}
where $d_m$ represents the overlap between the original message $s_i$
and the estimated message $\widehat{s}_i$ and is defined as
\begin{eqnarray}
 d_m &=& \frac{1}{K}\sum_{i=1}^{K} s_i \widehat{s}_i
  \label{eqn:def_overlap_m}
\end{eqnarray}
Image quality is usually measured by peak signal-to-noise ratio (PSNR).
However, since we deal with binary images, the image quality can also be
measured by $\mathrm{BER}$ as
\begin{eqnarray}
 \mathrm{BER}_R &=& \frac{1-d_R}{2}, 
\end{eqnarray}
where the overlap, $d_R$, between
the original image $f_{\mu}$ and the estimated image $\widehat{f}_{\mu}$
is defined as
\begin{eqnarray}
 d_R &=& \frac{1}{N}\sum_{\mu=1}^{N} f_{\mu} \widehat{f}_{\mu}.
\end{eqnarray}
Because mean squared error $\mathrm{MSE}=4\,\mathrm{BER}_R$, PSNR can be
calculated from $\mathrm{BER}_R$.

Using the bit error rate (BER), we evaluate the performance of our
proposed method, which estimates both messages and image at the same time.
We want to know the average performance rather than specific messages
and image. Therefore, we average the BER over all possible messages
$\boldsymbol{s}$, images $\boldsymbol{f}$, and spread codes
$\xi_i^{\mu}$. We assume random diffusion by spread codes and a large
system limit.
Under this assumptions, we can derive saddle point equations of overlaps
$m$ and $R$ from the posterior probability and can then theoretically 
evaluate the performance. 

\subsection{Replica method}

In order to determine the average performance, Helmholtz free energy $F$
is averaged over messages, the pixel value of images, and spread codes.
That is, $[F] = -T [\log Z] $,
where $[\cdot]$ denotes a configurational average defined by
\begin{eqnarray}
 \left[x\right] 
 &=& \int\prod_{\mu=1}^{N}dr_{\mu} 
  \mathrel{\mathop{\mathrm{Tr}}\limits_{\small \boldsymbol{s},\boldsymbol{f}}}
  \left<P\left(\boldsymbol{r}\vert\boldsymbol{s},\boldsymbol{f}\right) 
   P\left(\boldsymbol{s}\right)P\left(\boldsymbol{f}\right) x
\right>_{\xi} ,
\end{eqnarray}
and $\left<\cdot\right>_{\xi}$ denotes an average over $\xi_i^{\mu}$.
By using the replica method, we can obtain this averaged free energy $[F]$
from the relation
\begin{eqnarray}
 [\log Z] &=& \lim_{n\to0} \frac{[Z^n]-1}{n}. 
  \label{eqn:logZ2Zn}
\end{eqnarray}
In other words, $[\log Z]$ can be calculated from $n$ replicas of the
original system using the configurational average of the product of the 
partition functions, $Z^n$. We therefore start to calculate from
\begin{eqnarray}
 \left[Z^n\right] 
  &=& \int\prod_{\mu=1}^{N}dr_{\mu} 
  \left< 
    \mathrel{\mathop{\mathrm{Tr}}\limits_{\small \boldsymbol{s},\boldsymbol{f}}}
    P\left(\boldsymbol{r} \vert \boldsymbol{s},\boldsymbol{f}\right) 
    P\left(\boldsymbol{s}\right) P\left(\boldsymbol{f}\right)
    Z^n
  \right>_{\xi}  \label{eqn:Zn} \\
  &=& 
  \int\prod_{\mu=1}^{N}dr_{\mu} \Trs{\boldsymbol{s},\boldsymbol{x}^a} 
  \Trs{\boldsymbol{f}, \boldsymbol{g}^a} 
  \left< \left(2\pi\sigma_0^2\right)^{-\frac{N}{2}}
    \exp\left[-\frac{1}{2\sigma_0^2}\sum_{\mu=1}^N 
      \left(r_{\mu}-\frac{1}{\sqrt{K}}\sum_{i=1}^{K} \xi^{\mu}_is_i
        -f_{\mu} \right)^2 
      \right. \right. \nonumber \\ && 
      -\frac{1}{2\sigma^2}\sum_{a=1}^n \sum_{\mu=1}^N 
      \left(r_{\mu}-\frac{1}{\sqrt{K}} 
        \sum_{i=1}^{K} \xi^{\mu}_i x_i^{a}-g_{\mu}^{a}\right)^2
      \left. \left.
      +\frac{\alpha_0}{N}\sum_{\mu<\nu}f_{\mu}f_{\nu}
      +\frac{\alpha}{N}\sum_{a=1}^n \sum_{\mu<\nu}
      g_{\mu}^ag_{\nu}^a
    \right] \right>_{\xi}, \label{eqn:def_Zn}
\end{eqnarray}
where $a$ is the replica index.

According to the replica analysis of the CDMA model
\cite{TTanaka2001,Nishimori2001}, we first need to carry out the terms
of the messages.  Let us average over the spread codes $\xi_i^{\mu}$. By
introducing the following notations to (\ref{eqn:def_Zn}):
\begin{eqnarray}
 v^{\mu}_0 = \frac{1}{\sqrt{K}}\sum_{i=1}^{K} \xi^{\mu}_i s_i &,\;\;&
 v^{\mu}_a = \frac{1}{\sqrt{K}}\sum_{i=1}^{K} \xi^{\mu}_i x_i^{a} ,
\end{eqnarray}
we obtain 
\begin{eqnarray}
 \left[Z^n\right] &=& \int dv^{\mu}_0 \prod_{a} dv^{\mu}_a e^{N(g_1+g_2)} , \\
  e^{Ng_1} &=& \Trs{\boldsymbol{s},\boldsymbol{x}} \prod_{\mu}
  \left\langle 
  \delta\left(v^{\mu}_0-\frac{1}{\sqrt{K}}\sum_{i=1}^{K} \xi^{\mu}_i s_i\right)
  \prod_{a}\delta\left(v^{\mu}_a\!-\!\frac{1}{\sqrt{K}}\sum_{i=1}^{K} 
                     \xi^{\mu}_i x_i^{a}\right) 
    \right\rangle_{\xi} , \\
e^{Ng_2} &=& \Trs{\boldsymbol{f},\boldsymbol{g}}\prod_{\mu} 
  \int\frac{dr_{\mu}}{\sqrt{2\pi}\sigma_0}
  \exp\left[-\frac{1}{2\sigma_0^2}\left(r_{\mu}-v^{\mu}_0-f_{\mu}\right)^2 
       -\frac{1}{2\sigma^2}\sum_{a}\left(r_{\mu}-v^{\mu}_{a}-g^a_{\mu}\right)^2
       \right. \nonumber \\ && \left. 
     +\frac{\alpha_0}{N}\sum_{\mu<\nu}f_{\mu}f_{\nu}
       +\frac{\alpha}{N}\sum_{a=1}^n \sum_{\mu<\nu}
       g_{\mu}^ag_{\nu}^a
     \right] .
\end{eqnarray}
In the term $e^{Ng_1}$, using the integral representation of delta
function $\delta(\cdot)$, we can carry out the average over the spread codes
$\xi_i^{\mu}$ and then introduce order parameters to the terms of
the messages $s_i, x_i^a$, given by
\begin{eqnarray}
 q_{ab} &=& \frac{1}{K}\sum_{i=1}^K x_i^a x_i^b ,\;\;
 m_{a} = \frac{1}{K}\sum_{i=1}^K s_i x_i^a .
\end{eqnarray}
The term $e^{Ng_1}$ can be represented as
\begin{eqnarray}
 e^{Ng_1} &=& \Trs{\boldsymbol{s},\boldsymbol{x}} 
  \left\{\prod_{a<b}\int dq_{ab}
   \delta\left(Kq_{ab}-\sum_{i=1}^K x_i^a x_i^b\right)
   \prod_{a}\int dm_{a}\delta\left(Km_{a}-\sum_{i=1}^K s_i x_i^a\right)
  \right\} 
  \nonumber\label{181804_10Apr12} \\ &\times&
   \prod_{\mu}
  \int \frac{d\widehat{v}^{\mu}_0}{2\pi} \prod_{a} 
  \frac{d\widehat{v}^{\mu}_a}{2\pi} 
  \exp\left[i\widehat{v}^{\mu}_0v^{\mu}_0
       +i\sum_{a}\widehat{v}^{\mu}_av^{\mu}_a
       -\frac{1}{2}(\widehat{v}^{\mu}_0)^2
       -\frac{1}{2}\sum_{a}(\widehat{v}^{\mu}_a)^2
       \right. \nonumber \\ && \left. 
       -\sum_{a<b}q_{ab} \widehat{v}^{\mu}_a\widehat{v}^{\mu}_b
       -\sum_{a}m_{a} \widehat{v}^{\mu}_0\widehat{v}^{\mu}_a
         \right] \\ %
 &=& \int \prod_{a<b}\frac{idq_{ab}d\widehat{q}_{ab}}{2\pi} 
  \prod_{a}\frac{idm_{a}d\widehat{m}_{a}}{2\pi} 
  \exp\left[ -K\sum_{a<b}\widehat{q}_{ab}q_{ab}
       -K\sum_{a}\widehat{m}_{a}m_a \right] \nonumber \\ && 
  \times \prod_{\mu}\int \frac{d\widehat{v}^{\mu}_0}{2\pi} 
  \prod_{a}\frac{d\widehat{v}^{\mu}_a}{2\pi}
  \exp\left[i\widehat{v}^{\mu}_0v^{\mu}_0+i\sum_{a}\widehat{v}^{\mu}_av^{\mu}_a
       -\frac{1}{2}\sum_{a}(\widehat{v}^{\mu}_a)^2
       -\sum_{a<b}q_{ab} \widehat{v}^{\mu}_a\widehat{v}^{\mu}_b
       \right. \nonumber \\ && \left.
       -\sum_{a}m_{a} \widehat{v}^{\mu}_0\widehat{v}^{\mu}_a
       -\frac{1}{2}(\widehat{v}^{\mu}_0)^2 \right] \nonumber \\ && 
  \times \Trs{\boldsymbol{s},\boldsymbol{x}} 
  \prod_{k=1}^K \exp\left[\sum_{a<b}\widehat{q}_{ab} x_k^a x_k^b
		     +\sum_{a}\widehat{m}_{a} s_k x_k^a \right] .
  \label{eqn:msg_ma_qab}
\end{eqnarray}

From integrating $\left[Z^n\right]$ into the terms of $r_{\mu},
v_0^{\mu}$, and $\widehat{v}^{\mu}_0$, we can obtain the term
(\ref{eqn:apn_eG3_pre}). (See appendix~\ref{apd:Zn} for more details on
this derivation.)  Now, we assume symmetry between replicas for the order
parameters of messages; that is, $q_{ab} = q, \;
\widehat{q}_{ab}=\widehat{q}, \; m_{a} = m$, and $ \widehat{m}_a =
\widehat{m}$.  Under this assumption, we obtain
\begin{eqnarray}
\left[Z^n\right] &=& \int \frac{idqd\widehat{q}}{2\pi} 
 \frac{idmd\widehat{m}}{2\pi} e^{N(G_1+G_2+G_3)} , \label{eqn:ZnG} \\
 G_1 &=& -\frac{1}{2}n(n-1)\beta\widehat{q}q-n\beta\widehat{m}m , \\
 G_2 &=& -\frac{n\beta\widehat{q}}{2}
  +n\beta\int D_z\log2\cosh\left(z\sqrt{\widehat{q}}+\widehat{m}\right) , \\
 e^{NG_3} &=&  \Trs{\boldsymbol{f},\boldsymbol{g}} \prod_{\mu}
  \int \frac{dv^{\mu}_0d\widehat{v}^{\mu}_0}{2\pi} 
  \prod_{a}\frac{dv^{\mu}_ad\widehat{v}^{\mu}_a}{2\pi} 
  \frac{dr_{\mu}}{\sqrt{2\pi}\sigma_0}
  \exp\left[i\widehat{v}^{\mu}_0v^{\mu}_0+i\sum_{a}\widehat{v}^{\mu}_av^{\mu}_a
   -\frac{1}{2}\sum_{a}(\widehat{v}^{\mu}_a)^2
   -\frac{1}{2}(\widehat{v}^{\mu}_0)^2 \right. \nonumber \\ 
  && \left.
      -q\sum_{a<b} \widehat{v}_a\widehat{v}_b
      -m\sum_{a} \widehat{v}_0\widehat{v}_a
      -\frac{1}{2\sigma_0^2} \left(r_{\mu}-v^{\mu}_0-f_{\mu}\right)^2 
      -\frac{1}{2\sigma^2}\sum_{a} \left(r_{\mu}-v^{\mu}_a-g_{\mu}^{a} \right)^2
      \right.  \nonumber \\ 
 && 
  \left. +\frac{\alpha_0}{N}\sum_{\mu<\nu}f_{\mu}f_{\nu}
   +\frac{\alpha}{N}\sum_{a=1}^n \sum_{\mu<\nu} g_{\mu}^ag_{\nu}^a \right] ,
\end{eqnarray}
where $D_z=dz/\sqrt{2\pi}e^{-z^2/2}$.
By integrating over $v^{\mu}_a, \widehat{v}^{\mu}_a$, the term
$e^{NG_3}$ is given by
\begin{eqnarray}
 e^{NG_3} &=& \Trs{\boldsymbol{f},\boldsymbol{g}} \prod_{\mu}
 \sigma^{n}
 \left(\sigma^2+1-q\right)^{-\frac{n}{2}} 
 \left\{1+\frac{n(2m-q-\sigma_0^2-1)}{2(\sigma^2+1-q)}
 +\frac{n(\sigma_0^2+1)}{2\sigma^2} 
    +\Upsilon \left(\sum_ag_{\mu}^a\right)^2
    \right\}
 \nonumber \\ 
 && \times \exp\left[\Phi +\Psi\sum_{a<b}g_{\mu}^ag_{\mu}^b 
	 +\Omega f_{\mu}\sum_{a}g_{\mu}^a
	 +\frac{\alpha_0}{2N}\left(\sum_{\mu=1}^Nf_{\mu}\right)^2
	 +\frac{\alpha}{2N}\sum_a\left(\sum_{\mu=1}^Ng_{\mu}^a\right)^2
	       \right]. 
 \label{eqn:eG3_va_copy}
\end{eqnarray}
(See appendix~\ref{apn:eG3} for more details on this derivation.)
This term represents contribution from the image.

Next, for term $e^{NG_3}$, we introduce various order parameters of 
the images, given by
\begin{eqnarray}
 r_0 &=& \frac{1}{N}\sum_{\mu=1}^N f_{\mu}  ,\;\;
  r_a = \frac{1}{N}\sum_{\mu=1}^N g_{\mu}^a , \label{eqn:def_r0_ra} \\
 R_a &=& \frac{1}{N}\sum_{\mu=1}^N f_{\mu}g_{\mu}^a ,\;\; 
  Q_{ab} = \frac{1}{N}\sum_{\mu=1}^N g_{\mu}^ag_{\mu}^b .
  \label{eqn:def_Ra_Qab}
\end{eqnarray}
Using these order parameters, we can rewrite it as 
\begin{eqnarray}
 e^{NG_3} &=& \int dr_0 \prod_adr_a \prod_adR_a \prod_{a<b}dQ_{ab}
  e^{N(G_4+G_5+G_6+G_7)} , 
\end{eqnarray}
where
\begin{eqnarray}
  e^{NG_4} &=& \Trs{\boldsymbol{f}} \Trs{\boldsymbol{g}} 
  \exp\left[
       -\widehat{r}_0\left(Nr_0-\sum_{\mu=1}^{N}f_{\mu}\right)
       -\sum_a\widehat{r}_a\left(Nr_a-\sum_{\mu=1}^{N}g_{\mu}^a\right)
                  \right.  \nonumber \\
 && \left. 
     -\sum_a\widehat{R}_a\left(NR_a-\sum_{\mu=1}^{N}f_{\mu}g_{\mu}^a\right)
     -\sum_{a<b}\widehat{Q}_{ab}\left(NQ_{ab}-\sum_{\mu=1}^{N}g_{\mu}^ag_{\mu}^b\right)  
   \right] , \\
 e^{NG_5} &=& \left[ \sigma^{n}\left(\sigma^2+1-q\right)^{-\frac{n}{2}} 
              \right]^{N} , \\
 e^{NG_6} &=& \exp N
  \left[\frac{n(2m-q-\sigma_0^2-1)}{2(\sigma^2+1-q)}
   +\frac{n(\sigma_0^2+1)}{2\sigma^2} 
   +\Upsilon\left(2\sum_{a<b}Q_{ab}+n \right) \right] , \\
 e^{NG_7} &=& \exp N\left[ \Phi + 
      \Psi \sum_{a<b}Q_{ab}  
      +\Omega \sum_a R_a 
      +\frac{\alpha_0}{2}r_0^2 +\frac{\alpha}{2}\sum_ar_a^2
       \right] .
\end{eqnarray}
The variables $\Upsilon, \Phi, \Psi$, and $\Omega$ are given by
(\ref{eqn:Upsilon})--(\ref{eqn:Omega}).
We assume the replica symmetry for these order parameters; that is,
$r_a = r ,\; \widehat{r}_a = \widehat{r} ,\;
 R_a = R ,\;  \widehat{R}_a = \widehat{R} , 
 Q_{ab} = Q$, and $\widehat{Q}_{ab} = \widehat{Q}$.
They lead to 
\begin{eqnarray}
e^{NG_3} &=& \int \frac{idr_0d\widehat{r}_0}{2\pi}
 \int\frac{idrd\widehat{r}}{2\pi}
 \int\frac{idRd\widehat{R}}{2\pi}
 \int \frac{idQd\widehat{Q}}{2\pi}
 e^{N(G_4+G_5+G_6+G_7)} ,
\end{eqnarray}
where
\begin{eqnarray}
 G_4 &=& -r_0\widehat{r}_0 -nr\widehat{r} -nR\widehat{R} 
  -\frac{n(n-1)}{2} Q\widehat{Q} -\frac{n}{2}\widehat{Q}
  \nonumber \\
 && +\log\left\{2\cosh\left(\widehat{r}_0\right)
	  +n\Trs{f} \exp\left(\widehat{r}_0f\right)
	  \int D_s\log2\cosh\left(s\sqrt{\widehat{Q}}+\widehat{r}+\widehat{R}f\right)
	 \right\} , \\
 G_5 &=& n\log\sigma -\frac{n}{2}\frac{\sigma_0^2+1}{\sigma^2}
  -\frac{n}{2}\log\left(\sigma^2+1-q\right) , \\
 G_6 &=& \frac{n(2m-q-\sigma_0^2-1)}{2(\sigma^2+1-q)}
   +\frac{n(\sigma_0^2+1)}{2\sigma^2} 
   \nonumber \\
 && +n\left\{-\frac{2m-q-\sigma_0^2-1}{2(\sigma^2+1-q)^2}
      +\frac{(\sigma_0^2+1)\left(1-\frac{nm}{\sigma^2}\right)}{\sigma^2(\sigma^2+1-q)} 
      +\frac{\sigma_0^2+1}{2\sigma^4}
     \right\}\left(1-Q+nQ\right) , \\
 G_7 &=& \Phi + \frac{n(n-1)}{2}\Psi Q  
      +n\Omega R 
      +\frac{\alpha_0}{2}r_0^2 +\frac{n\alpha}{2}r^2 .
\end{eqnarray}

Thus, we obtain
\begin{eqnarray}
 \left[Z^n\right] &=& 
  \int\frac{idqd\widehat{q}}{2\pi} \frac{idm d\widehat{m}}{2\pi}
  \frac{idr_0d\widehat{r}_0}{2\pi} \frac{idrd\widehat{r}}{2\pi} 
  \frac{idRd\widehat{R}}{2\pi}     \frac{idQd\widehat{Q}}{2\pi}
  e^{N\left(G_1+G_2+G_4+G_5+G_6+G_7\right)} .
\end{eqnarray}
In the large-system limit $N\to\infty$, the integral can be evaluated
by the saddle point method.
From (\ref{eqn:logZ2Zn}), the free energy $F$ is given in the limit
$n\to0$ as 
\begin{eqnarray}
 F &=& \frac{1}{2}\beta\widehat{q}q-\beta\widehat{m}m
  -\frac{\beta\widehat{q}}{2}
  +\beta\int D_z\log2\cosh\left(z\sqrt{\widehat{q}}+\widehat{m}\right)
  +\log\sigma 
  -\frac{1}{2}\log\left(\sigma^2+1-q\right) 
 \nonumber \\  && 
 +\frac{2m-q-\sigma_0^2-1}{2(\sigma^2+1-q)} 
+\frac{\alpha}{2}r^2 -r\widehat{r} -R\widehat{R} 
 -\frac{1}{2} \left(1-Q\right) \widehat{Q}
  -\left(1-Q\right)\frac{2m-q-\sigma_0^2-1}{2(\sigma^2+1-q)^2}
 \nonumber \\
 &&   -\frac{1-R}{\sigma^2+1-q}
 +\frac{\Trs{f} \exp\left(\widehat{r}_0f\right)
  \int D_s\log2\cosh\left(s\sqrt{\widehat{Q}}+\widehat{r}+\widehat{R}f\right)}
  {2\cosh\left(\widehat{r}_0\right)} . 
\end{eqnarray}
Since there are $n$-independent constant terms in $F$, we define them as
\begin{eqnarray}
 F_0 &=& -r_0\widehat{r}_0 +\frac{\alpha_0}{2}r_0^2
  +\log2\cosh(\widehat{r}_0) .
\end{eqnarray}
Extremization of the free energy yields the saddle point equations as
\begin{eqnarray}
  m &=& \int D_z\tanh\left(z\sqrt{\widehat{q}}+\widehat{m}\right), \\
 \widehat{m} &=& \frac{1}{\beta(\sigma^2+1-q)} 
 -\frac{1-Q}{\beta(\sigma^2+1-q)^2} , \\
 q &=& \int D_z\tanh^2\left(z\sqrt{\widehat{q}}+\widehat{m}\right) , \\
 \widehat{q} &=& \frac{q-2m+\sigma_0^2+2(1-R)+Q}{\beta(\sigma^2+1-q)^2}
 -2(1-Q)\frac{q-2m+\sigma_0^2+1}{\beta(\sigma^2+1-q)^3} ,
  \\
 r &=&  \frac{1}{2\cosh(\widehat{r}_0)} \!
  \Trs{f} \!\! e^{\widehat{r}_0f} \!\! \int \!\! D_z
  \tanh\left(\alpha r+z\sqrt{\widehat{Q}}+\widehat{R}f\right) , 
  \nonumber \\ && \\
 R &=&  \frac{1}{2\cosh(\widehat{r}_0)} \!
  \Trs{f} \!\! fe^{\widehat{r}_0f} \!\! \int \!\! D_z
  \tanh\left(\alpha r+z\sqrt{\widehat{Q}}+\widehat{R}f\right) , 
  \nonumber \\ && \\
Q &=& \frac{1}{2\cosh(\widehat{r}_0)} \!
  \Trs{f} \!\! e^{\widehat{r}_0f} \!\! \int \!\! D_z
  \tanh^2\left(\alpha r+z\sqrt{\widehat{Q}}+\widehat{R}f\right) ,
  \nonumber \\ && \\
 \widehat{R} &=& \frac{1}{\sigma^2+1-q} ,\;\;\;  
 \widehat{Q} = \frac{q-2m+\sigma_0^2+1}{(\sigma^2+1-q)^2} , \\ 
 r_0 &=& \tanh(\widehat{r}_0) , \;\; \widehat{r}_0 = \alpha_0r_0 . 
  \label{eqn:r0=tanh}
\end{eqnarray}
In these equations, we can find two sets of equations for both the CDMA
model \cite{TTanaka2001,TTanaka2002,Nishimori2001} and the image
restoration model \cite{Nishimori2001}. These two equations depend on
each other.

\section{computer simulations} \label{sec:sim}

\subsection{Overlaps and BER}

Let us derive the overlaps $d_m$ and $d_R$. The overlaps are
averaged over all realization of the spreading codes and noises
\cite{Nishimori2001,TTanaka2002}. Therefore, the overlaps are given by
\begin{eqnarray}
 d_m &=& \lim_{n\to0} \lim_{K\to\infty} 
  \left[\frac{1}{K}\sum_{i=1}^{K} s_i 
	 \mathrm{sgn}\left(\left<\widehat{s}_i\right>_{\sigma}\right)\right], \\
 d_R &=& \lim_{n\to0} \lim_{N\to\infty} 
  \left[\frac{1}{N}\sum_{\mu=1}^{N} f_{\mu} 
	 \mathrm{sgn}\left(\left<\widehat{f}_{\mu}\right>_{\sigma}\right)\right], 
\end{eqnarray} 
where $\left<\cdot\right>_{\sigma}$ denotes the average over the
posterior distribution and $\left[\cdot\right]$ denotes the
average over the spreading codes, noises, messages and images
\cite{TTanaka2002}. We have
\begin{eqnarray}
 d_m &=& \int_{-\infty}^{\infty} Dz\; 
  \mathrm{sgn}\left(z\sqrt{\widehat{q}}+\widehat{m}\right) \\
 &=& \mathrm{erf}\left(\frac{\widehat{m}}{\sqrt{2\widehat{q}}}\right) , \\
 d_R &=& \frac{1}{2\cosh(\widehat{r}_0)} \!
 \Trs{f} \!\! fe^{\widehat{r}_0f} \!\! \int_{-\infty}^{\infty} \!\! D_z
 \mathrm{sgn}\left(\alpha r+z\sqrt{\widehat{Q}}+\widehat{R}f\right)  
 \\
 &=& \frac{1}{2\cosh(\widehat{r}_0)}
  \left\{e^{\widehat{r}_0}
   \mathrm{erf}\left(\frac{\alpha r+\widehat{R}}{\sqrt{2\widehat{Q}}}\right) 
  -e^{-\widehat{r}_0}
  \mathrm{erf}\left(\frac{\alpha r-\widehat{R}}{\sqrt{2\widehat{Q}}}\right)
  \right\},
\end{eqnarray}
where $\mathrm{erf}(x)$ is the error function defined by
\begin{eqnarray}
 \mathrm{erf}\left(x\right) &=& \frac{2}{\sqrt{\pi}}
   \int_0^{x}e^{-t^2}dt. 
\end{eqnarray}
From the overlaps, the BERs can be given by
\begin{eqnarray}
 \mathrm{BER}_m &=&
  \frac{1}{2}\left\{1-\mathrm{erf}\left(\frac{\widehat{m}}{\sqrt{2\widehat{q}}}
				  \right)\right\}  , \\
 \mathrm{BER}_R &=& \frac{1}{2}\left[1-\frac{1}{2\cosh(\widehat{r}_0)}
  \left\{e^{\widehat{r}_0}
   \mathrm{erf}\left(\frac{\alpha r+\widehat{R}}{\sqrt{2\widehat{Q}}}\right) 
  -e^{-\widehat{r}_0}
  \mathrm{erf}\left(\frac{\alpha r-\widehat{R}}{\sqrt{2\widehat{Q}}}\right)
  \right\}\right] .
\end{eqnarray}

\subsection{Verification of saddle point equations}

\begin{figure}[tb]
 \includegraphics[width=86mm]{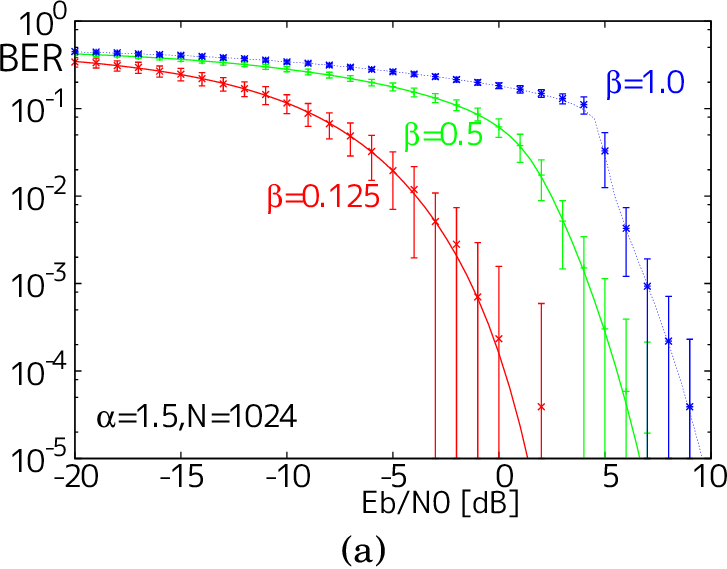}

 \includegraphics[width=86mm]{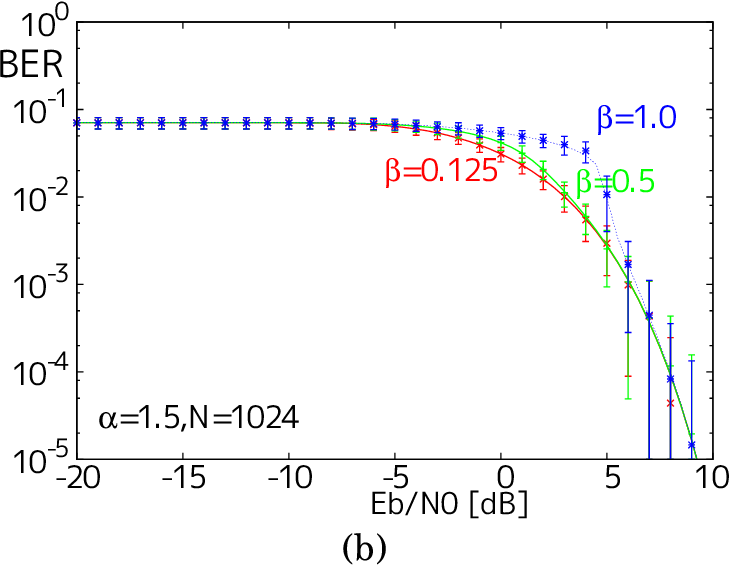}

 \caption{(a) Bit error rate $\mathrm{BER}_m$ for
 messages and (b) $\mathrm{BER}_R$ for image.
 The smooth parameter is $\alpha=1.5$. 
 The embedding rates are $\beta=0.125, 0.5, 1.0.$}
 \label{fig:SN_BER}
\end{figure}

We verify the obtained saddle point equations by computer simulations.
First, we consider the infinite range model for the image restoration
model. Figure~\ref{fig:IRMSample} shows the sample images generated 
that satisfy the prior probability (\ref{eqn:apriori_f}), where 
$\alpha_0=1.0, 1.5$, and $2.0$. In the case of (b) $\alpha_0=1.5$, the average
value of the pixels is $r_0=0.859$ from (\ref{eqn:r0=tanh}).
The size of sample images in Fig.~\ref{fig:IRMSample} is
$256\times256$ pixels. Since the length of the spread codes is
$N=1024$,
we use smaller original images of $32\times32$ pixels for the computer simulations. 
The message lengths are $K=128$, $512$ and $1024$. 
Figure~\ref{fig:SN_BER} shows the bit error rate ($\mathrm{BER}$) as a
function of the channel noise. The parameters in the decoder, $\alpha$
and $\sigma^2$, are given by true values $\alpha=\alpha_0$ and
$\sigma^2=\sigma_0^2$. The abscissa axis represents $E_b/N_0$ given by
\begin{equation}
 \frac{E_b}{N_0} = 10 \log_{10}\left(\frac{1}{2\sigma^2}\right) 
  [\mathrm{dB}], 
\end{equation}
where $\sigma^2$ is the variance of the Gaussian channel.  The axis of
ordinate represents the BERs for both messages $\mathrm{BER}_m$ and
images $\mathrm{BER}_R$. 
$\mathrm{BER}_m$ is averaged over $200$ trials in the computer
simulations.
The average $\mathrm{BER}_m$ is shown with error bars.
$\mathrm{BER}_R$ is calculated on whole image
and is shown with points.
The initial values of the estimated messages and estimated image are set
by the true values, and then we obtain one of the best solutions.  The
theoretical values obtained by the saddle point equations are plotted by
a solid line for embedding rate 
$\beta=K/N=0.125$ $(K=128)$, a dashed line for $\beta=0.5$ $(K=512)$, 
and a double-dashed line for $\beta=1.0$ $(K=1024)$.
The computer simulations results agreed with those derived theoretically.
In Fig.~\ref{fig:SN_BER}, the $\mathrm{BER}_m$ for the messages 
worsened according to the embedding rate $\beta$, while the
$\mathrm{BER}_R$ for the images were slightly influenced by $\beta$
under the fixed smooth parameter $\alpha$.

\begin{figure}[tb]
 \includegraphics[width=86mm]{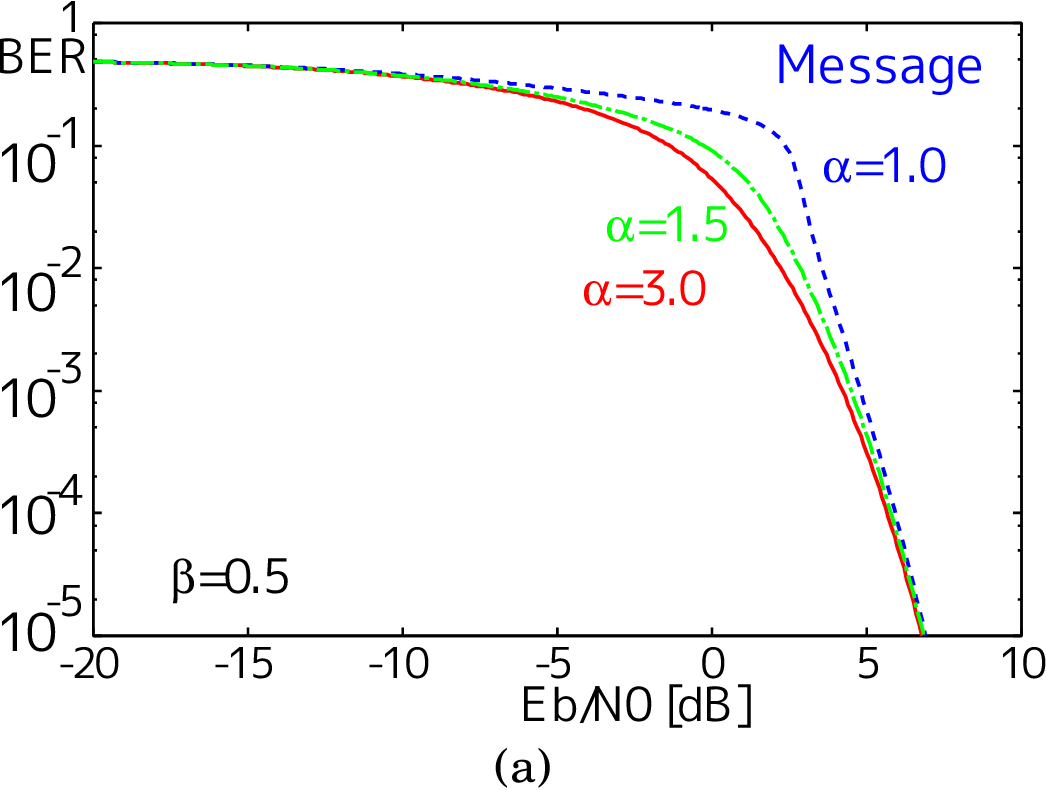}

 \includegraphics[width=86mm]{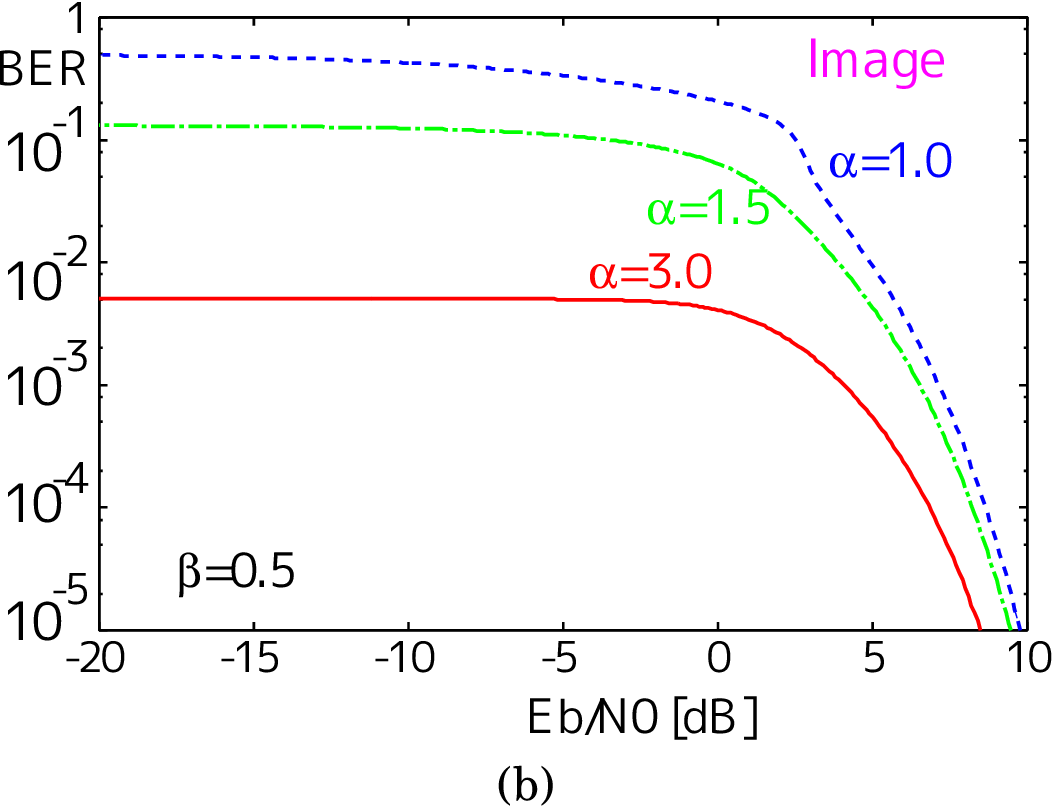}
 \caption{(a) Bit error rate $\mathrm{BER}_m$ for messages
 and (b) $\mathrm{BER}_R$ for image. 
 The smooth parameters are $\alpha=1.0, 1.5, 3.0$.
 The embedding rate is $\beta=0.5$. }
 \label{fig:BER_IRM_beta0.5}
\end{figure}

Next, we evaluate the bit error rate for the smooth parameters
$\alpha=1.0, 1.5$, and $3.0$ under the fixed embedding rate $\beta$.
Figure~\ref{fig:BER_IRM_beta0.5} shows the BERs for the embedding rate
$\beta=0.5$. Because of the fixed embedding rate, the $\mathrm{BER}_m$
for the messages were slightly influenced by the parameter $\alpha$,
while the $\mathrm{BER}_R$ for the images became better according to
$\alpha$. In other words, smoother images can be easily restored.

\subsection{Advantages of image restoration}

The key concept underlying the proposed method is that it can estimate
both the messages and image at the same time in the decoder. Here, we
compare the performance of the blind decoder with that of the informed
decoder.  Cases in which the original image is known or informed to the
decoder correspond to the CDMA model, and only messages are estimated.

\begin{figure}[tb]
 \includegraphics[width=86mm]{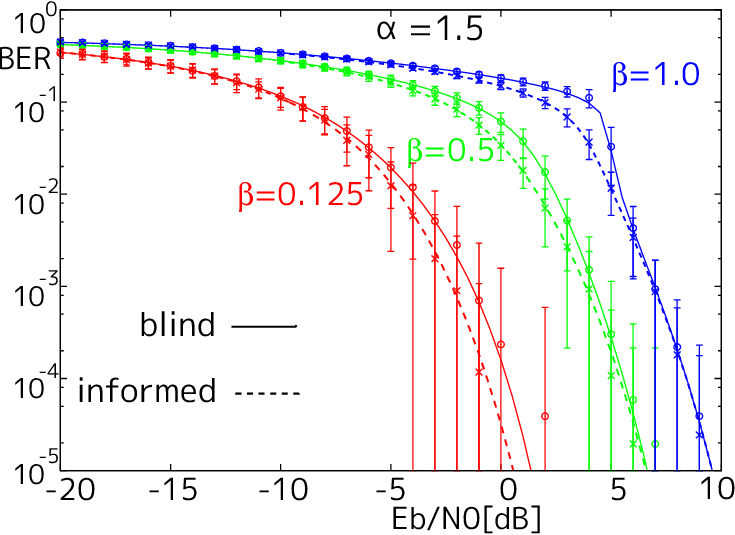}
 \caption{$\mathrm{BER}_m$ in the cases of blind and 
 informed images.}
 \label{fig:WMvsCDMA}
\end{figure}

Figure~\ref{fig:WMvsCDMA} shows the bit error rate $\mathrm{BER}_m$ for
messages in the blind and informed decoders.  The embedding rates are
$\beta=0.125, 0.5$, and $1.0$. The BERs in the blind decoder are larger
than those in the informed decoder because images are also
estimated. However, in cases in which the embedding rate $\beta$ is
small enough, or in which there is not much noise in the communication
channel, there is not much difference between the blind and informed
decoders, i.e., blind decoder can successfully carry out
image estimation.

\subsection{2D Ising Model}

\begin{figure}[tb]
 \hfill\includegraphics[width=50mm]{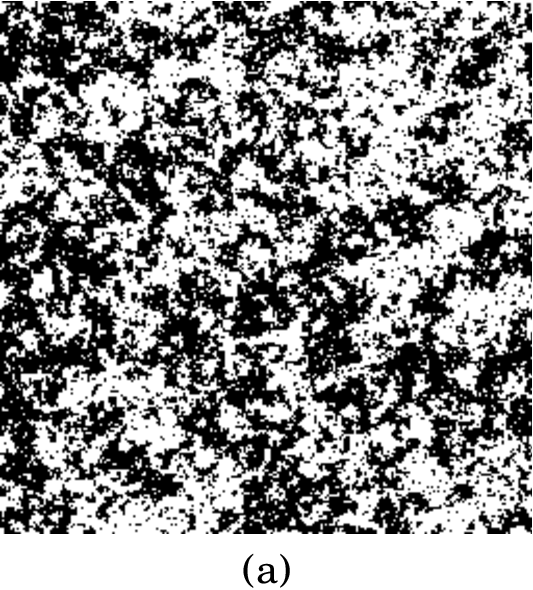}
 \hfill\includegraphics[width=50mm]{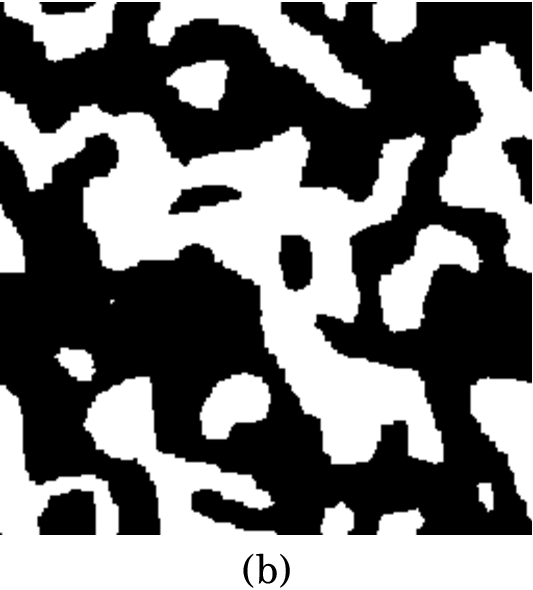}
 \hfill\includegraphics[width=50mm]{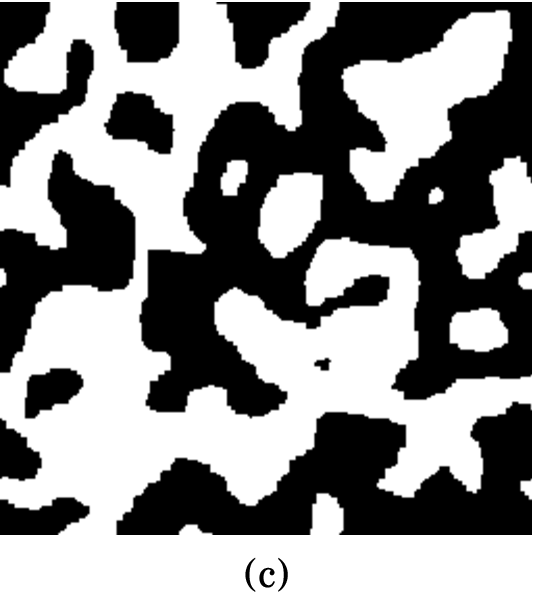}
 \hfill\mbox{}
 \caption{Images generated by 2D-Ising model ($256\times256$ pixels)
 with the smooth parameters (a)~$\alpha_0=0.4$, (b)~$\alpha_0=1.5$, and
 (c) $\alpha_0=10.0$.}
 \label{fig:2DIsingSample}
\end{figure}

In addition to the infinite range model, we also consider the 2D Ising
model for image restoration, in which each pixel is statically connected
with four-neighbors.
This model is natural for the image restoration.
 In this model, there are some clusters in
generated images because the pixels interact with their nearest
neighbors.  These cluster patterns can be seen in the parity of JPEG
images.  In this section, we treat the 2D Ising model as an image
generating model; that is, the prior probability is given by
\begin{eqnarray}
  P\left(\boldsymbol{f}\right) 
  &\propto& \exp\left[\alpha_0
                 \sum_{\left<\mu,\nu\right>}f_{\mu}f_{\nu}\right], 
  \label{eqn:apriori_2DIsing}
\end{eqnarray}
where $\left<\mu,\nu\right>$ denotes pairs of nearest neighbor sites. 
Figure~\ref{fig:2DIsingSample} shows the generated images for 
parameters $\alpha_0=0.4, 1.5$, and $10$ in the 2D Ising model.
In this manner, once the generating models have been changed,
the generated images are much different. Since it is difficult to
construct a generating model of natural images, it is necessary
to consider various generating models in which as many characteristics of 
natural images are applied as possible.

\begin{figure}[tb]
 \includegraphics[width=100mm]{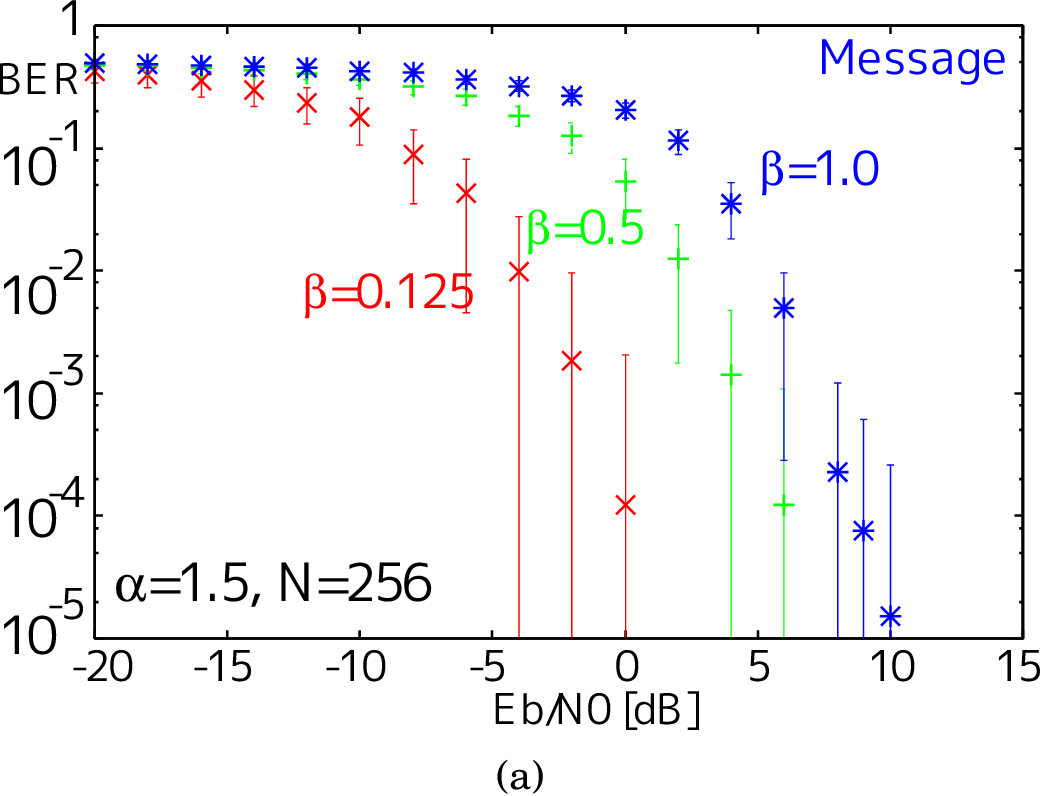} 

 \includegraphics[width=100mm]{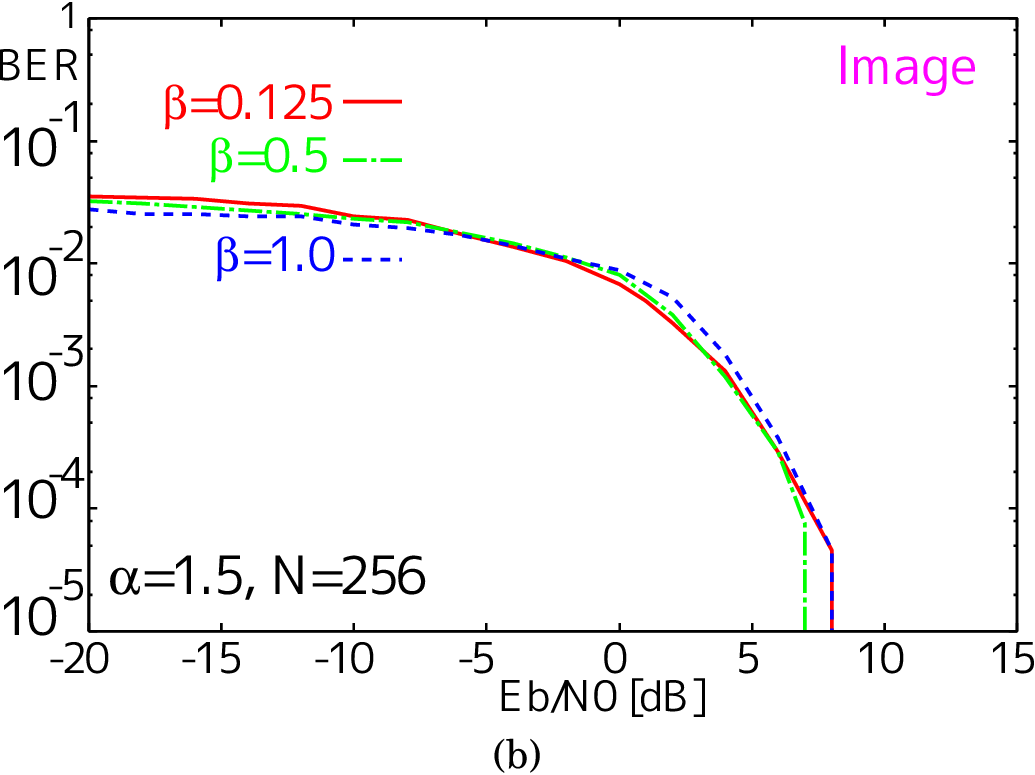} 

 \caption{(a) Bit error rates $\mathrm{BER}_m$ for messages and 
 (b) $\mathrm{BER}_R$ for image in 2D Ising model. 
 The smooth parameter is $\alpha=\alpha_0=1.5$.
 The embedding rates are $\beta=0.125, 0.5, 1.0$.}
 \label{fig:SN_BER_2DI}
\end{figure}

The posterior probability of the original image $\boldsymbol{f}$
given the tampered image $\boldsymbol{r}$ is given by
\begin{eqnarray}
 P\left(\boldsymbol{f}\vert \boldsymbol{r}\right) 
 &=& \frac{1}{Z} 
  \exp\left[-\frac{1}{2\sigma^2}
   \sum_{\mu=1}^N \left(r_{\mu}-f_{\mu}\right)^2  
   +\alpha\sum_{\left<\mu,\nu\right>}f_{\mu}f_{\nu} \right] .
   \label{eqn:post_f_2D}
\end{eqnarray}
Although the replica method can be applied to a certain 2D Ising model
with diluted random connections by using a mean field approximation 
\cite{EdwardsAnderson1975},
the exact treatment of the 2D Ising model is technically difficult
and the replica method does not yield the accurate assessment.
We therefore evaluate its performance by computer simulations.
Since we can see the continuous structure in Fig.~\ref{fig:2DIsingSample}, 
the image size in the 2D Ising model is $256\times256$ pixels unlike
ones of the infinite range model. The images are divided into $256$
blocks, whose size is $256$ pixels per a block.
So, the spread code length is $N=256$. 
Figure~\ref{fig:SN_BER_2DI} shows the bit error rates $\mathrm{BER}_m$
and $\mathrm{BER}_R$ for the 2D Ising model. The parameters $\alpha$
and $\sigma^2$ are set to the true value $\alpha=\alpha_0$ and
$\sigma^2=\sigma_0^2$. 
The BERs are averaged over all blocks.
$\mathrm{BER}_R$ for the images are slightly influenced by the embedding
rate $\beta$ under the fixed parameter $\alpha$.

\begin{figure}[tb]
\begin{center}
 \includegraphics[width=100mm]{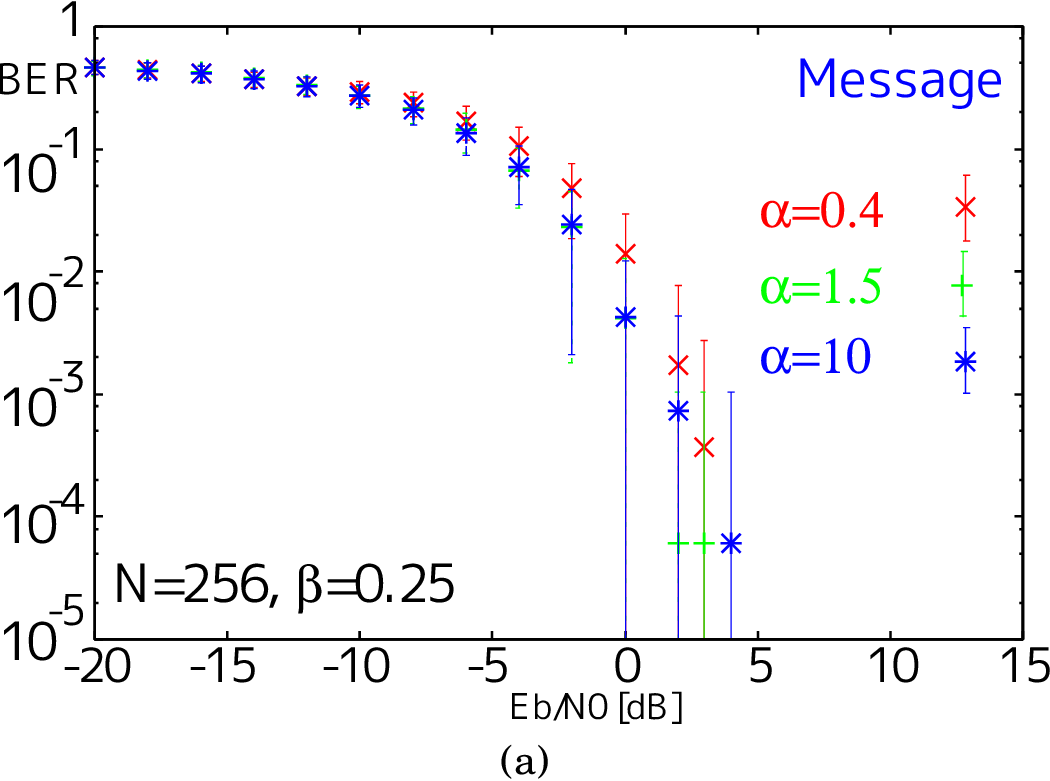} 

 \includegraphics[width=100mm]{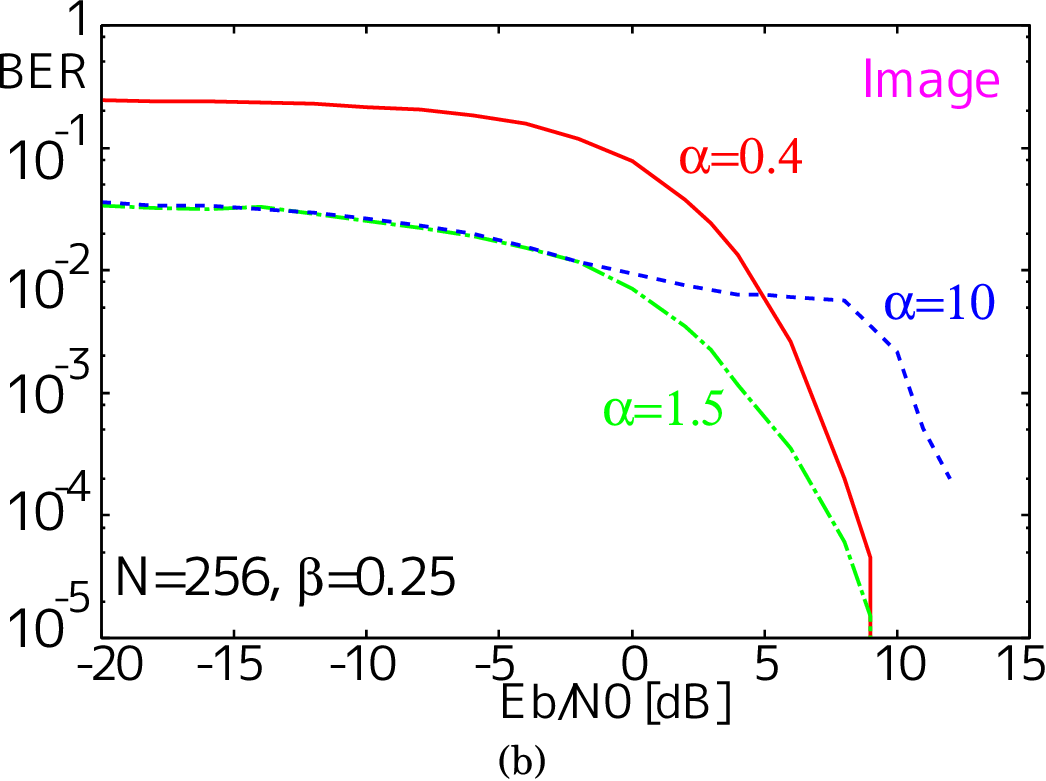} 
\end{center}
 \caption{(a) Bit error rate $\mathrm{BER}_m$ for messages and 
 (b) $\mathrm{BER}_R$ for image in 2D Ising model. 
 The smooth parameters are $\alpha_0=0.4, 1.5$, $10.0$.
 The embedding rate is $\beta=0.25$. }
 \label{fig:SN_BER_beta}
\end{figure}

Next, we evaluate the performance under the fixed embedding rate
$\beta=0.25$. Figure~\ref{fig:SN_BER_beta} shows BERs for the smooth
parameters $\alpha_0=0.4, 1.5$, and $10.0$. $\mathrm{BER}_m$ for messages are
slightly influenced by $\alpha$. $\mathrm{BER}_R$ for images
in cases of $\alpha_0>1$ are smaller than those of $\alpha=0.4$.
For large $\alpha_0=10$, phase transition may occure. 

\begin{figure}[tb]
 \includegraphics[width=100mm]{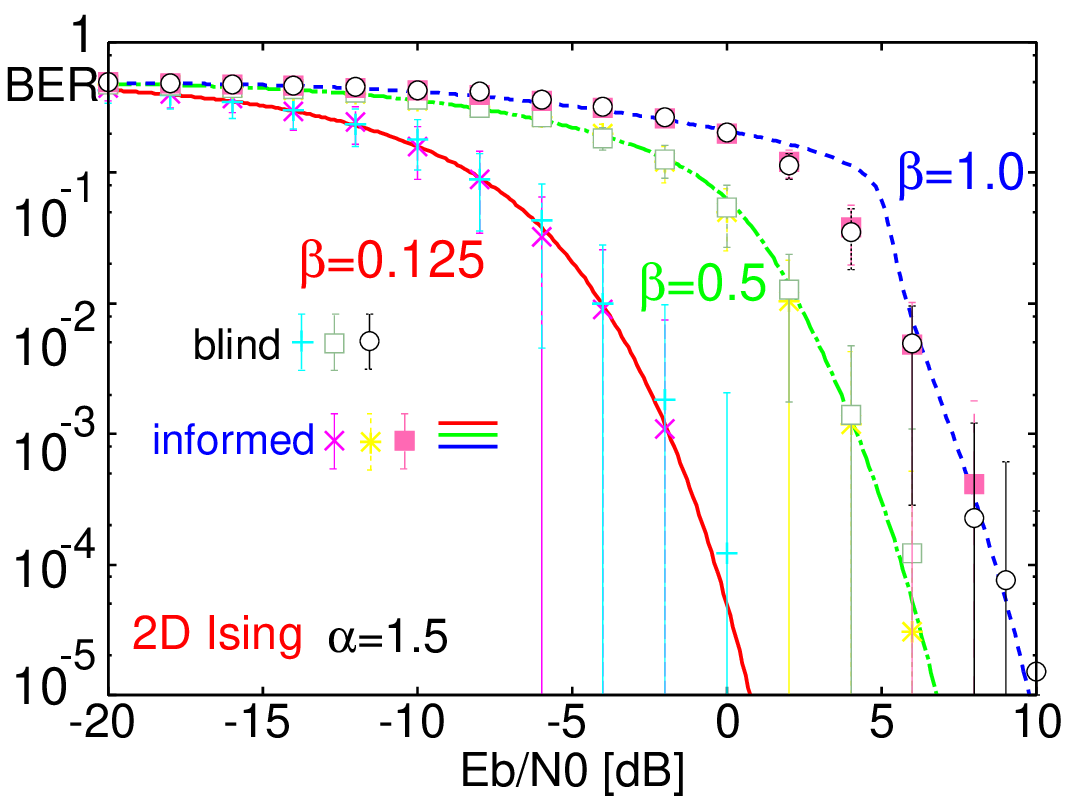}
 \caption{$\mathrm{BER}_m$ in blind and informed cases.
 The smooth parameter is $\alpha=1.5$. The embedding rates are
 $\beta=0.125, 0.5$, $1.0$. }
 \label{fig:WMvsCDMA_2D}
\end{figure}

Figure~\ref{fig:WMvsCDMA_2D} shows the bit error rate $\mathrm{BER}_m$
for messages in both blind and informed decoders. The embedding rates were
$\beta=0.125, 0.5$, and $1.0$. Curved lines denote the theoretical values for
the informed decoder. The blind decoder had just as good a performance as 
the informed decoder. That is, the blind decoder could successfully restore
the image and estimate the messages.

\section{Conclusion} \label{sec:conc}

We proposed an estimation method that can estimate messages and an image
at the same time when using a blind decoder. When this method is used
with Bayes estimation, prior probabilities for both the messages and the
images are required. In this paper, we assumed that the prior probability
for messages had a uniform distribution and that those for 
images were the infinite range model and 2D Ising model.

For the infinite range model, we derived the saddle point equations by
the replica method in order to evaluate the average performance.  Since
there are two terms -- the messages term and the image term -- 
we implemented a two-step approach: first, we introduced
order parameters for the messages and assumed replica symmetry
for them, and second, we introduced order parameters for the image and 
assumed replica symmetry for them. The
obtained saddle point equations consist of two indivisible parts: the
equations of the CDMA model and those of the image restoration model.
We verified the saddle point equations by computer simulations.  The
theoretical results agreed with those of the simulations.

Next, we evaluated the performance of the 2D Ising model by computer
simulations. When the smooth parameter $\alpha_0$ was fixed, there was
little change in the bit error rate (BER) for images, and the BER
for messages depended on the embedding rate $\beta$.  In contrast, when
the embedding rate $\beta$ was fixed, there was little change in the BER
for messages, and the BER for images depended on
the smooth parameter. However, there was a lower bound in the 2D Ising
model.

We also evaluated the performance differences between blind and informed
decoders. Results showed that the difference was very small when the
embedding or attack rates were small, since the image restoration could
still be carried out well. This demonstrates the effectiveness of the
proposed method.

\begin{acknowledgments}
This work was supported by JSPS KAKENHI Grant Numbers 21700255,
 JP16K00156, JP16K05474.
The computer simulations were carried out on PC clusters at Yamaguchi
University and on multi-core processors at Nara Women's University.
\end{acknowledgments}

\input{WM_Replica_arXiv.bbl}


\appendix
\section{Integral of $[Z^n]$ with respect to $r_{\mu}, v_0^{\mu}, \widehat{v}^{\mu}_0$}
\label{apd:Zn}

From (\ref{eqn:msg_ma_qab}), we obtain
\begin{eqnarray}
 \left[Z^n\right]
  &=&\int \prod_{a<b}\frac{idq_{ab}d\widehat{q}_{ab}}{2\pi} 
 \prod_{a}\frac{idm_{a}d\widehat{m}_{a}}{2\pi}
 e^{N(G_1+G_2+G_3)} , \label{eqn:ZnG3}
\end{eqnarray}
where 
\begin{eqnarray}
 e^{G_1} &=& 
  \exp\left[-\beta\sum_{a<b}\widehat{q}_{ab}q_{ab}
       -\beta\sum_{a}\widehat{m}_{a}m_a\right] , \\ 
 e^{NG_2} &=& \Trs{\boldsymbol{s},\boldsymbol{x}} 
  \prod_{k=1}^K \exp\left[\sum_{a<b}\widehat{q}_{ab} x_k^a x_k^b
		     +\sum_{a}\widehat{m}_{a} s_k x_k^a \right] , \\ 
 e^{NG_3} &=&  \Trs{\boldsymbol{f},\boldsymbol{g}} \prod_{\mu}
  \int \frac{dv^{\mu}_0d\widehat{v}^{\mu}_0}{2\pi} 
  \prod_{a}\frac{dv^{\mu}_ad\widehat{v}^{\mu}_a}{2\pi} 
  \frac{dr_{\mu}}{\sqrt{2\pi}\sigma_0}
  \exp\left[i\widehat{v}^{\mu}_0v^{\mu}_0+i\sum_{a}\widehat{v}^{\mu}_av^{\mu}_a
   -\frac{1}{2}\sum_{a}(\widehat{v}^{\mu}_a)^2
   -\frac{1}{2}(\widehat{v}^{\mu}_0)^2 \right. \nonumber \\ 
  && \left.
      -\sum_{a<b}q_{ab} \widehat{v}_a\widehat{v}_b
      -\sum_{a}m_{a} \widehat{v}_0\widehat{v}_a
      -\frac{1}{2\sigma_0^2} \left(r_{\mu}-v^{\mu}_0-f_{\mu}\right)^2 
      -\frac{1}{2\sigma^2}\sum_{a} \left(r_{\mu}-v^{\mu}_a-g_{\mu}^{a} \right)^2
      \right. 
    \nonumber \\ 
 && 
  \left.
	+\frac{\alpha_0}{N}\sum_{\mu<\nu}f_{\mu}f_{\nu}
	+\frac{\alpha}{N}\sum_{a=1}^n \sum_{\mu<\nu}
	g_{\mu}^ag_{\nu}^a
    \right] ,
\end{eqnarray}
Now, we integrate $e^{NG_3}$ by $r_{\mu}, v_0^{\mu}, \widehat{v}^{\mu}_0$:
\begin{eqnarray}
 \lefteqn{e^{NG_3} }\nonumber \\
  &=&  \Trs{\boldsymbol{f},\boldsymbol{g}} \prod_{\mu}
  \sqrt{\frac{\sigma^2}{\sigma^2+n(\sigma_0^2+1)}} 
  \int\prod_{a}\frac{dv^{\mu}_ad\widehat{v}^{\mu}_a}{2\pi} 
  \int D_{t_{\mu}} 
  \exp\left[ 
       -\frac{1}{2\sigma^2}\sum_{a}(v^{\mu}_a)^2
       +i\sum_{a}\widehat{v}^{\mu}_av^{\mu}_a 
       -\frac{1}{2}\sum_{a}(\widehat{v}^{\mu}_a)^2
      \right. \nonumber \\ 
  && +\left\{
  t_{\mu}\sqrt{\frac{\sigma_0^2+1}{\sigma^2\{\sigma^2+n(\sigma_0^2+1)\}}}
   +\frac{1}{\sigma^2+n(\sigma_0^2+1)}
  \left(-i\sum_{a}m_{a}\widehat{v}^{\mu}_a +f_{\mu}
   +\frac{\sigma_0^2+1}{\sigma^2}\sum_{a}g^{a}_{\mu} \right)
      \right\} \sum_{a}v^{\mu}_a 
  \nonumber \\
 && -\frac{1}{\sigma^2}\sum_{a}g_{\mu}^{a}v^{\mu}_a 
   +\frac{n}{2\{\sigma^2+n(\sigma_0^2+1)\}}
  \left(\sum_{a}m_{a}\widehat{v}^{\mu}_a\right)^2 
  +\frac{in}{\sigma^2+n(\sigma_0^2+1)}
  f_{\mu}\sum_{a}m_{a}\widehat{v}^{\mu}_a
\nonumber \\ 
 && \left. +\frac{1}{\sigma^2+n(\sigma_0^2+1)}
  \left(-i\sum_{a}m_{a}\widehat{v}^{\mu}_a+f_{\mu}\right)
  \sum_{a}g^{a}_{\mu} 
  +\frac{\sigma_0^2+1}{2\sigma^2\{\sigma^2+n(\sigma_0^2+1)\}}
  \left(\sum_{a}g^{a}_{\mu}\right)^2
  \right. \nonumber \\ && \left.
  -\sum_{a<b}q_{ab} \widehat{v}^{\mu}_a\widehat{v}^{\mu}_b
  +\frac{\alpha_0}{N}\sum_{\mu<\nu}f_{\mu}f_{\nu}
  +\frac{\alpha}{N}\sum_{a=1}^n \sum_{\mu<\nu}
  g_{\mu}^ag_{\nu}^a
			  \right] . 
\label{eqn:apn_eG3_pre}
\end{eqnarray}

Under the assumption of the replica symmetry, we obtain
\begin{eqnarray}
  \lefteqn{e^{NG_3} }\nonumber \\
  &=&  \Trs{\boldsymbol{f},\boldsymbol{g}} \prod_{\mu}
  \sqrt{\frac{\sigma^2}{\sigma^2+n(\sigma_0^2+1)}} 
  \int\prod_{a}\frac{dv^{\mu}_ad\widehat{v}^{\mu}_a}{2\pi} 
  \int D_{t_{\mu}} 
  \exp\left[ 
       -\frac{1}{2\sigma^2}\sum_{a}(v^{\mu}_a)^2
       +i\sum_{a}\widehat{v}^{\mu}_av^{\mu}_a 
       -\frac{1}{2}\sum_{a}(\widehat{v}^{\mu}_a)^2
      \right. \nonumber \\ 
  && +\left\{
  t_{\mu}\sqrt{\frac{\sigma_0^2+1}{\sigma^2\{\sigma^2+n(\sigma_0^2+1)\}}}
   +\frac{1}{\sigma^2+n(\sigma_0^2+1)}
  \left(-im \sum_{a}\widehat{v}^{\mu}_a +f_{\mu}
   +\frac{\sigma_0^2+1}{\sigma^2}\sum_{a}g^{a}_{\mu} \right)
      \right\} \sum_{a}v^{\mu}_a 
  \nonumber \\
 && -\frac{1}{\sigma^2}\sum_{a}g_{\mu}^{a}v^{\mu}_a 
   +\frac{nm^2}{2\{\sigma^2+n(\sigma_0^2+1)\}}
   \left(\sum_{a}\widehat{v}^{\mu}_a\right)^2 
  +\frac{inm}{\sigma^2+n(\sigma_0^2+1)}
  f_{\mu}\sum_{a}\widehat{v}^{\mu}_a
\nonumber\label{184006_10Apr12} \\ 
 && \left. +\frac{1}{\sigma^2+n(\sigma_0^2+1)}
  \left(-im\sum_{a}\widehat{v}^{\mu}_a+f_{\mu}\right)
  \sum_{a}g^{a}_{\mu} 
  +\frac{\sigma_0^2+1}{2\sigma^2\{\sigma^2+n(\sigma_0^2+1)\}}
  \left(\sum_{a}g^{a}_{\mu}\right)^2
  \right. \nonumber \\ && \left.
  -\frac{q}{2}\left\{\left(\sum_a\widehat{v}^{\mu}_a\right)^2
	      -\sum_a\left(\widehat{v}^{\mu}_a\right)^2\right\}
  +\frac{\alpha_0}{2N}\left(\sum_{\mu=1}^N f_{\mu}\right)^2
  -\frac{\alpha_0}{2}
  +\frac{\alpha}{2N}\sum_{a=1}^n 
  \left(\sum_{\mu=1}^N g_{\mu}^a\right)^2  -\frac{\alpha}{2} \right] . 
\label{eqn:apn_eG3_pre_RS}
\end{eqnarray}

\section{Integral of $e^{NG_3}$ with respect to $v^{\mu}_a,\widehat{v}^{\mu}_a$}
\label{apn:eG3}

Under the assumption of the replica symmetry,  we integrate by
$v^{\mu}_a$, and eliminate the terms at the limit $n\to0$. We obtain
\begin{eqnarray}
 \lefteqn{e^{NG_3} =  \Trs{\boldsymbol{f},\boldsymbol{g}} \prod_{\mu}
  \sqrt{\frac{\sigma^2}{\sigma^2+n(\sigma_0^2+1)}} 
  \int\prod_{a}\frac{dv^{\mu}_ad\widehat{v}^{\mu}_a}{2\pi} 
  \int D_{t_{\mu}} 
  \exp\left[ 
       -\frac{1}{2\sigma^2}\sum_{a}(v^{\mu}_a)^2
       +i\sum_{a}\widehat{v}^{\mu}_av^{\mu}_a 
      \right. } \nonumber \\ 
  && +\left\{
  t_{\mu}\sqrt{\frac{\sigma_0^2+1}{\sigma^2\{\sigma^2+n(\sigma_0^2+1)\}}}
   +\frac{1}{\sigma^2+n(\sigma_0^2+1)}
  \left(-im\sum_{a}\widehat{v}^{\mu}_a +f_{\mu}
   +\frac{\sigma_0^2+1}{\sigma^2}\sum_{a}g^{a}_{\mu} \right)
      \right\} \sum_{a}v^{\mu}_a 
  \nonumber \\
  && -\frac{1}{\sigma^2}\sum_{a}g_{\mu}^{a}v^{\mu}_a 
  -\frac{1}{2}\left(1-q\right)\sum_{a}(\widehat{v}^{\mu}_a)^2
  +\left\{\frac{nm^2}{2\{\sigma^2+n(\sigma_0^2+1)\}} -\frac{q}{2}
   \right\}
  \left(\sum_{a}\widehat{v}^{\mu}_a\right)^2 
  \nonumber \\ 
 && +\frac{im}{\sigma^2+n(\sigma_0^2+1)}
 \left(nf_{\mu}-\sum_{a}g^{a}_{\mu}\right)\sum_{a}\widehat{v}^{\mu}_a
 +\frac{1}{\sigma^2+n(\sigma_0^2+1)} f_{\mu}\sum_{a}g^{a}_{\mu} 
 \nonumber\label{eqn:eG3_n} \\ 
 && \left. 
  +\frac{\sigma_0^2+1}{2\sigma^2\{\sigma^2+n(\sigma_0^2+1)\}}
  \left(\sum_{a}g^{a}_{\mu}\right)^2
  +\frac{\alpha_0}{2N}\left(\sum_{\mu=1}^N f_{\mu}\right)^2
  +\frac{\alpha}{2N}\sum_{a=1}^n 
  \left(\sum_{\mu=1}^N g_{\mu}^a\right)^2 
			      \right] , 
    \\ 
 &=& \Trs{\boldsymbol{f},\boldsymbol{g}} \prod_{\mu}
 \sigma^{n}
  \int\prod_{a}\frac{d\widehat{v}^{\mu}_a}{\sqrt{2\pi}} \int D_{t_{\mu}} 
  \exp\left[ -\frac{1}{2}\left(\sigma^2+1-q\right)
       \sum_a (\widehat{v}^{\mu}_a)^2 
       +\left(m-\frac{q}{2} \right)
       \left(\sum_{a}\widehat{v}^{\mu}_a\right)^2 
      \right. \nonumber \\ 
  && +\frac{inm}{\sigma^2} f_{\mu}\sum_{a}\widehat{v}^{\mu}_a  
 -i\sum_a g_{\mu}^a\widehat{v}^{\mu}_a 
 +i\left(1-\frac{nm}{\sigma^2} \right)
  \left(f_{\mu} +\frac{\sigma_0^2+1}{\sigma^2}\sum_bg_{\mu}^b 
  +t_{\mu}\sqrt{\sigma_0^2+1}\right) 
  \sum_a \widehat{v}^{\mu}_a  
  \nonumber \\
 && 
  +\frac{n}{2\sigma^2} 
  \left(f_{\mu}+\frac{\sigma_0^2+1}{\sigma^2}\sum_a g_{\mu}^a \right)^2 
  -\frac{\sigma_0^2+1}{2\sigma^4}
  \left(\sum_{a}g^{a}_{\mu}\right)^2 
  \nonumber \\
 && +\frac{nt_{\mu}\sqrt{\sigma_0^2+1}}{\sigma^2}
  \left(f_{\mu}+\frac{\sigma_0^2+1}{\sigma^2}\sum_a g_{\mu}^a \right) 
  -\frac{t_{\mu}\sqrt{\sigma_0^2+1}}{\sigma^2} \sum_a g_{\mu}^a 
  +\frac{nt_{\mu}^2(\sigma_0^2+1)}{2\sigma^2} 
  \nonumber \\
 && \left.  
     +\frac{\alpha_0}{2N}\left(\sum_{\mu=1}^N f_{\mu}\right)^2
     +\frac{\alpha}{2N}\sum_{a=1}^n 
     \left(\sum_{\mu=1}^N g_{\mu}^a\right)^2 
   \right] . 
\end{eqnarray}
Using Hubbard-Stratonovich transformation,
\begin{eqnarray}
 \exp\left[\left(m-\frac{q}{2} \right)
  \left(\sum_a \widehat{v}^{\mu}_a\right)^2\right]
 &=& \int D_{z_{\mu}}\exp
 \left[z_{\mu}\sqrt{2m-q}
  \sum_a \widehat{v}^{\mu}_a\right],
\end{eqnarray}
we integrate by $\widehat{v}^{\mu}_a, t_{\mu}, z_{\mu}$: 
\begin{eqnarray}
 e^{NG_3} &=& \Trs{\boldsymbol{f},\boldsymbol{g}} \prod_{\mu}
 \sigma^{n}
 \left(\sigma^2+1-q\right)^{-\frac{n}{2}} 
 \left\{1+\frac{n(2m-q-\sigma_0^2-1)}{2(\sigma^2+1-q)}
 +\frac{n(\sigma_0^2+1)}{2\sigma^2} 
    +\Upsilon \left(\sum_ag_{\mu}^a\right)^2
    \right\}
 \nonumber \\ 
 && \times \exp\left[\Phi +\Psi\sum_{a<b}g_{\mu}^ag_{\mu}^b 
	 +\Omega f_{\mu}\sum_{a}g_{\mu}^a
	 +\frac{\alpha_0}{2N}\left(\sum_{\mu=1}^Nf_{\mu}\right)^2
	 +\frac{\alpha}{2N}\sum_a\left(\sum_{\mu=1}^Ng_{\mu}^a\right)^2
	       \right], 
 \label{eqn:eG3_va}
\end{eqnarray}
where
\begin{eqnarray}
  \Upsilon &=& \frac{1}{2}
  \left\{
   -\frac{2m-q-\sigma_0^2-1}{(\sigma^2+1-q)^2} 
   +\frac{2(\sigma_0^2+1)\left(1-\frac{nm}{\sigma^2}\right)}
   {\sigma^2(\sigma^2+1-q)}
   +\frac{\sigma_0^2+1}{\sigma^4}
  \right\} , \label{eqn:Upsilon} \\
 \Phi &=& n\left\{\frac{\sigma_0^2+1}{\sigma^2(\sigma^2+1-q)}
  \left(1-\frac{nm}{\sigma^2}\right)
  -\frac{\sigma_0^2+1}{2\sigma^4} 
 -\frac{1}{\sigma^2+1-q} \right\} , \\ 
 \Psi &=& \frac{\sigma_0^2+1}{\sigma^2(\sigma^2+1-q)}
  \left(1-\frac{nm}{\sigma^2}\right)
  \left\{2-\frac{n(\sigma_0^2+1)}{\sigma^2}\left(1-\frac{nm}{\sigma^2}\right)
  \right\}  
  -\frac{\sigma_0^2+1}{\sigma^4} , \\
 \Omega &=& \frac{1}{\sigma^2+1-q}
  \left\{1 -n(\sigma_0^2+1)
   \left(1-\frac{nm}{\sigma^2}\right)^2 \right\} 
  +\frac{n(\sigma_0^2+1)}{\sigma^4} . \label{eqn:Omega}
\end{eqnarray}

\end{document}